\documentclass[aps,prl,reprint,superscriptaddress,floatfix,longbibliography]{revtex4-2}
\usepackage{blindtext}
\usepackage{graphicx} 
\usepackage{xcolor}
\usepackage{physics}
\usepackage{amsmath,amssymb,amsfonts}
\usepackage{braket}
\usepackage[utf8]{inputenc}
\usepackage[T1]{fontenc}
\usepackage{dcolumn}
\usepackage{makecell}
\usepackage{upgreek}
\usepackage{bm}

\usepackage[colorlinks=true,linkcolor=blue,citecolor=blue,urlcolor=blue]{hyperref}

\newcommand{\Lphys}{\mathfrak{L}_{\rm{phys}}}
\newcommand{\LL}{\mathfrak{L}_{L}}

\newcommand{\Vundet}{V_{\rm{u}}}
\newcommand{\Vdet}{V_{\rm{d}}}

\newcommand{\Pst}{\mathbb{P}}
\newcommand{\faulttraj}{\mathbb{Q}_{\bm{a}}}
\newcommand{\probphys}{p_{\rm{phys}}}
\newcommand{\probL}{p_{L}}

\newcommand{\GammaEDPEC}{\Gamma_{\rm ED+PEC}}

\newcommand{\Lundet}{\mathfrak{L}_{\rm u}}
\newcommand{\Ldet}{\mathfrak{L}_{\rm d}}

\begin{document}

\title{Spacetime mitigation of logical errors}

\author{Laurin~E.~Fischer}
\affiliation{IBM Research, R{\"u}schlikon, Switzerland}

\author{Ali Javadi-Abhari}
\affiliation{IBM Research, Yorktown Heights, NY}

\author{Simon Martiel}
\affiliation{IBM Research, Saclay, France}

\author{Alireza Seif}
\email{alireza.seif@ibm.com}
\affiliation{IBM Research, Yorktown Heights, NY}

\begin{abstract}
Error detection can suppress noise at a lower sampling cost than probabilistic error cancellation (PEC), but leaves residual logical errors. We combine the two approaches to mitigate these errors while lowering the cost of PEC. We introduce a spacetime Pauli-Lindblad representation and use it to construct the post-selected logical noise perturbatively from physical noise and syndrome information. Individually undetected faults contribute at first order, while detected faults with canceling syndromes generate higher-order terms coupling distinct spacetime locations. To reconcile the nonlinearity of post-selection with the linearity of PEC, we formulate the combined protocol in terms of unnormalized post-selected maps, with normalization performed only after averaging. On the \textit{ibm\_aachen} superconducting quantum processor, Clifford experiments validate the post-selected noise model and its perturbative scaling. We then demonstrate error detection with first-order PEC for transverse-field Ising dynamics using 22 data and 27 check qubits. The combined protocol recovers the mean magnetization within statistical uncertainty through six Trotter steps and reduces the inferred sampling overhead by up to a factor of 63 relative to PEC without post-selection.
\end{abstract}
\maketitle

\textit{Introduction---}
Before large-scale fault-tolerant quantum computation becomes available, error detection (ED) and error mitigation offer complementary ways to improve computational accuracy~\cite{QuantumErrorMitigation}. Error-detecting checks can suppress noise at a lower sampling cost than error mitigation through probabilistic error cancellation~\cite{Temme_PEC, endo_PEC} (PEC) by rejecting runs with a nontrivial syndrome, but leave residual logical errors that bias the post-selected results~\cite{martielLowoverheadErrorDetection2025,javadi-abhariBigCatsEntanglement2025,liaoAchievingComputationalGains2025}. PEC can mitigate modeled noise-induced bias~\cite{bergProbabilisticErrorCancellation2023,aharonovReliableHighaccuracyError2026, chen2026disambiguating}, but its sampling cost generally grows exponentially with circuit error strength. Applying PEC after error detection can therefore mitigate the residual logical errors at a lower sampling cost than PEC alone, provided that the savings outweigh the cost of rejected samples and the additional noise introduced by the checks~\cite{aharonovSyndromeAwareMitigation2025,zhongCombiningErrorDetection2025a}.

Combining ED with PEC poses both characterization and implementation challenges. Post-selection correlates faults across circuit locations, even when the underlying physical noise is uncorrelated, because individually detectable faults can survive jointly when their syndromes cancel. Characterizing the resulting logical noise therefore requires combining a model of the physical noise with the circuit's syndrome information. A direct approach would construct the full physical noise channel and post-select it, but the size of this channel grows exponentially with the number of qubits, making direct inference of the post-selected logical noise impractical beyond small systems~\cite{zhongCombiningErrorDetection2025a}. Moreover, post-selection is generally nonlinear in the input state, whereas PEC relies on linear combinations of maps with signed weights. Independently post-selecting and normalizing each randomized circuit can therefore alter the ensemble that PEC is intended to implement.

\begin{figure}
    \centering    \includegraphics[width=1\columnwidth]{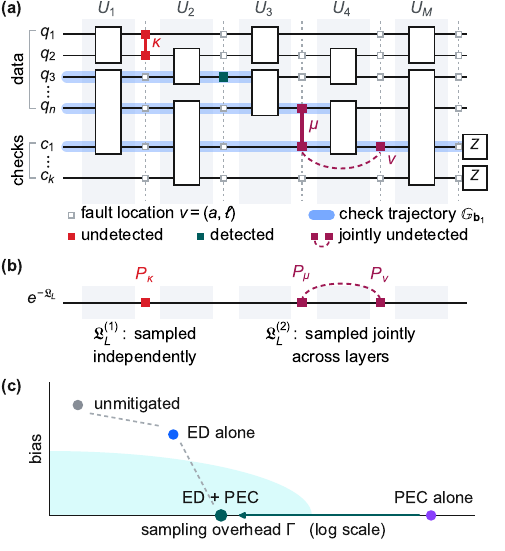}
    \caption{\textbf{Spacetime PEC with error detection.}
\textbf{a)} Circuit with $n$ data and $k$ check qubits. Blue highlights a check trajectory. Red faults evade detection;
green faults are detected; the magenta pair (dashed line) survives because
its syndromes cancel.
\textbf{b)} Spacetime PEC samples single-location Paulis at first
order and joint Paulis at second order.
\textbf{c)} Schematic bias--overhead tradeoff: ED+PEC can provide
unbiased mitigation with lower sampling overhead than PEC alone. }
    \label{fig:schematic}
\end{figure}

Here we introduce \emph{spacetime probabilistic error cancellation}, which combines ED and PEC to mitigate post-selected logical noise at reduced sampling cost. We develop a spacetime Pauli-Lindblad representation that incorporates a sparse physical noise model~\cite{bergProbabilisticErrorCancellation2023} and the circuit's syndrome information. Individually undetected faults contribute at first order, while detected faults with canceling syndromes generate higher-order correlations which span across space and time. This expansion provides a systematic construction of the inverse logical noise without requiring the full post-selected logical channel. To reconcile the nonlinearity of post-selection with the linearity of PEC, we formulate the combined protocol in terms of unnormalized post-selected maps and normalize only after averaging.

On \textit{ibm\_aachen}, a superconducting quantum processor with tunable couplers, we validate the post-selected logical noise model through second order using Clifford circuits and implement first-order spacetime PEC for non-Clifford Ising dynamics on 22 data and 27 check qubits. The combined protocol recovers the mean magnetization within statistical uncertainty through six Trotter steps and reduces the inferred sampling overhead, including the cost of post-selection, by up to a factor of 63 relative to PEC without post-selection.

\textit{Setup---}
We consider $N=n+k$ qubits, with $n$ data qubits and $k$ check qubits. The input state is $\rho=\rho_D\otimes\ket{0}\!\bra{0}^{\otimes k}$, where $\rho_D$ is a logical data input. The ideal circuit $U=U_M\cdots U_1$ implements the target computation on the data qubits while returning the checks to $\ket{0}^{\otimes k}$ for every logical data input. We model noise by Pauli faults $P_{a_\ell}\in\{I,X,Y,Z\}^{\otimes N}$ occurring after each circuit layer $\ell$. A sequence $\bm a=(a_1,\ldots,a_M)$ specifies a spacetime fault trajectory, with probability $\probphys(\bm a)$ and corresponding circuit operator $K_{\bm a}=P_{a_M}U_M\cdots P_{a_1}U_1$. The resulting noisy evolution is $
\mathcal C_{\rm phys}(\rho)=\sum_{\bm a}\probphys(\bm a)K_{\bm a}\rho K_{\bm a}^\dagger$~\cite{kam2026spatiotemporal}. For independent noise across layers, the trajectory distribution $\probphys(\bm a)$ factorizes over $\ell$, while correlations between layers lead to a nonfactorized distribution.

Terminal check measurements yield a syndrome $\bm s\in\{0,1\}^k$. We retain the trivial syndrome, with projector $\Pi_0=I_D\otimes\ket{0}\!\bra{0}^{\otimes k}$ and post-selected evolution
\begin{equation}
\mathcal C_L(\rho)=\frac{\Pi_0\mathcal C_{\rm phys}(\rho)\Pi_0}
{\operatorname{Tr}[\Pi_0\mathcal C_{\rm phys}(\rho)]}.
\label{eq:noisycircuitL}
\end{equation}
Our construction applies to circuits in which each Pauli fault trajectory has a deterministic syndrome $\bm s(\bm a)$ independent of the logical input state. Post-selection is then state independent, with $\alpha=\sum_{\bm a:\bm s(\bm a)=\bm0}\probphys(\bm a)$. The post-selected fault distribution is $\probL(\bm a)=\probphys(\bm a)\delta_{\bm s(\bm a),\bm0}/\alpha$, where $\delta$ is the Kronecker delta. Consequently, $\mathcal C_L(\rho)=\sum_{\bm a}\probL(\bm a)K_{\bm a}\rho K_{\bm a}^\dagger$ is linear on these inputs.

Pauli twirling or randomized compiling can tailor general noise into Pauli noise by averaging over randomized circuit implementations~\cite{wallmanNoiseTailoringScalable2016, hashimRandomizedCompilingScalable2021}. Combining these techniques with post-selection, however, introduces an additional challenge. Each post-selected expectation is normalized by the probability of passing the checks, which can depend on the input state and the circuit randomization. This makes post-selection generally nonlinear and can change the randomized ensemble. We resolve this by multiplying each post-selected expectation by its post-selection probability before averaging, thereby undoing its normalization. We then divide the average by the overall post-selection probability, yielding the linear evolution used for PEC (Appendix~\ref{app:linearized})~\cite{martiel2026sampling}.

Our goal is to invert the post-selected logical noise encoded by $\probL$. A sparse generator provides a tractable representation when a direct construction is too costly. Because post-selection can correlate faults across circuit layers, the resulting generator is generally nonlocal in time.

\textit{Spacetime Pauli-Lindblad noise---}
To represent faults at different circuit locations, we introduce an auxiliary Pauli space with one tensor factor per layer~\cite{xie2026noiseagnostic}. This space is used solely to keep track of fault locations. Let $V$ be the set of elementary faults in the characterized physical noise model. Each label $\nu=(a,\ell)\in V$ represents $P_a$ after layer $\ell$, with auxiliary operator $\Pst_\nu$ acting as $P_a$ on that layer and as the identity elsewhere. We use a sparse Pauli-Lindblad model~\cite{bergProbabilisticErrorCancellation2023} and, throughout this work, consider single-qubit and nearest-neighbor two-qubit Pauli faults independently at each layer, giving
\begin{equation}
    \Lphys(\cdot)
    =
    \sum_{\nu\in V}
    \lambda_\nu
    \left[
        \Pst_\nu(\cdot)\Pst_\nu-(\cdot)
    \right].
    \label{eq:phys-lind}
\end{equation}
A fault trajectory is represented by $\faulttraj=\bigotimes_{\ell=1}^M P_{a_\ell}$, with irrelevant Pauli phases omitted. The auxiliary channel $e^{\Lphys}(\cdot)=\sum_{\bm a}\probphys(\bm a)\faulttraj(\cdot)\faulttraj^\dagger$ encodes the trajectory distribution. Physical correlations between layers can be represented by generator terms spanning multiple layers.

For Clifford circuits, propagating an output check operator $Z$ backward to each fault location gives a sequence of Pauli operators. Their tensor product defines the check trajectory $\mathbb G_{\bm b_j}$, where $\bm b_j$ labels that sequence (Appendix~\ref{app:syndrome-rule}). The syndrome bit $s_{\nu,j}$ is zero when $\Pst_\nu$ commutes with $\mathbb G_{\bm b_j}$ and one when they anticommute~\cite{delfosse2023simulationnoisycliffordcircuits,delfosseSpacetimeCodesClifford2023}. The syndrome of a collection of faults is the bitwise sum $\oplus$ (addition modulo two) of their individual syndromes. This rule also applies to non-Clifford circuits when every propagated check commutes with each non-Clifford gate it encounters, so those gates leave the check trajectory unchanged.

The auxiliary logical channel is $e^{\LL}(\cdot)=\sum_{\bm a}\probL(\bm a)\faulttraj(\cdot)\faulttraj^\dagger$. We construct its generator $\LL$ by expanding the logarithm of the channel in the physical fault rates. The logarithm removes contributions generated by products of lower-order terms, leaving only the connected contributions, or cumulants, at each order. Partition $V=\Vundet\sqcup\Vdet$ into individually undetected faults ($\bm s_\nu=\bm0$) and detected faults ($\bm s_\nu\ne\bm0$). Because undetected faults do not affect post-selection, their contribution factors from the auxiliary channel exactly. The logarithm therefore separates their contribution to the generator, so no connected cumulant mixes detected and undetected faults (Appendix~\ref{app:higher-order}).

We express the generator as $\LL=\sum_{j\geq1}\LL^{(j)}$, where order $j$ denotes total degree $j$ in the physical rates. Individually undetected faults retain their original generator exactly
\begin{equation}
    \LL^{(1)}(\cdot)
    =
    \sum_{\nu\in\Vundet}
    \lambda_\nu
    \left[
        \Pst_\nu(\cdot)\Pst_\nu-(\cdot)
    \right].
    \label{eq:first_order_generator}
\end{equation}
The perturbative approximation therefore concerns only combinations of detected faults that survive post-selection. Their leading contribution comes from pairs whose syndromes cancel
\begin{equation}
    \LL^{(2)}(\cdot)
    =
    \frac{1}{2}
    \sum_{\substack{
        \nu,\mu\in\Vdet\\
        \bm{s}_\nu\oplus\bm{s}_\mu=\bm{0}
    }}
    \lambda_\nu\lambda_\mu
    \left[
        \Pst_\mu\Pst_\nu(\cdot)\Pst_\nu\Pst_\mu
        -
        (\cdot)
    \right],
    \label{eq:logical-generator-second-order-main}
\end{equation}
where the factor $1/2$ removes double counting. All terms of order $j\geq2$ arise solely from $\Vdet$ and describe connected contributions from combinations with zero total syndrome (Appendix~\ref{app:higher-order}).

\textit{Spacetime PEC---}
Once $\LL$ has been constructed to a chosen order, PEC is implemented by quasiprobabilistically sampling the inverse post-selected logical error channel, $e^{-\LL}$. Because higher-order generator terms couple faults at distinct circuit locations, the corresponding Pauli branches are sampled jointly across spacetime. For each retained term with coefficient $\kappa$, the identity and Pauli branches are sampled with the appropriate signs, giving a quasiprobability norm $e^{2\max(\kappa,0)}$ and hence no quasiprobability overhead for $\kappa<0$~\cite{seif2026single}. All sampled branches have zero total syndrome and preserve the post-selection probability. Measurement outcomes receive the corresponding PEC weights and the post-selection rescaling described above.

Let $\gamma_L$ denote the quasiprobability norm of the factorized inverse. The corresponding sampling-cost measure for error detection combined with PEC is $\Gamma_{\mathrm{ED+PEC}}=\gamma_L^2/\alpha$, where $\gamma_L^2$ accounts for quasiprobability sampling and $1/\alpha$ for post-selection. Its leading expansion is
\begin{equation}
    \Gamma_{\mathrm{ED+PEC}}
    =
    \frac{1}{\alpha}
    \exp\left[
        4\sum_{\nu\in\Vundet}\max(\lambda_\nu,0)
        +O(\lambda_d^2)
        \right],
    \label{eq:ed-pec-overhead}
\end{equation}
where $\lambda_d$ is the characteristic magnitude of the detected-fault rates and the remainder is second order at fixed circuit structure. For comparison, let $V_{\mathrm{bare}}$ denote the elementary faults in the same computation without error detection. The corresponding factorized PEC cost is
\begin{equation}
    \Gamma_{\mathrm{PEC}}
    =
    \exp\left[
        4\sum_{\nu\in V_{\mathrm{bare}}}
        \max(\lambda_\nu,0)
        \right].
    \label{eq:pec-overhead}
\end{equation}
Thus, detected faults are removed from the PEC exponent at first order and enter only through higher-order syndrome-canceling combinations. ED+PEC is advantageous when this reduction outweighs the cost of syndrome rejection and the additional noise introduced by the checks, hence the need for gate-efficient error detection.

The standard way of achieving error detection is via encoding in a distance-2 CSS code \cite{froland2026utility,zhongCombiningErrorDetection2025a}. Even though the space overhead of this approach is rather tame, this leads to a large increase in physical gate count, leading to poor overall circuit performances. Other approaches propose to exploit symmetries of space-inefficient encodings to detect violations and discard faulty runs \cite{papicNearTermFermionicSimulation2026}. Alternatively, one can also design \emph{ad hoc} error detection gadgets tailored to the target's architecture and thus achieving high fault coverage and a milder gate cost, leading to gate-efficient error detection \cite{martielLowoverheadErrorDetection2025,martiel2026sampling, javadi-abhariBigCatsEntanglement2025}. 

\begin{figure}
    \centering
    \includegraphics[width=\columnwidth]{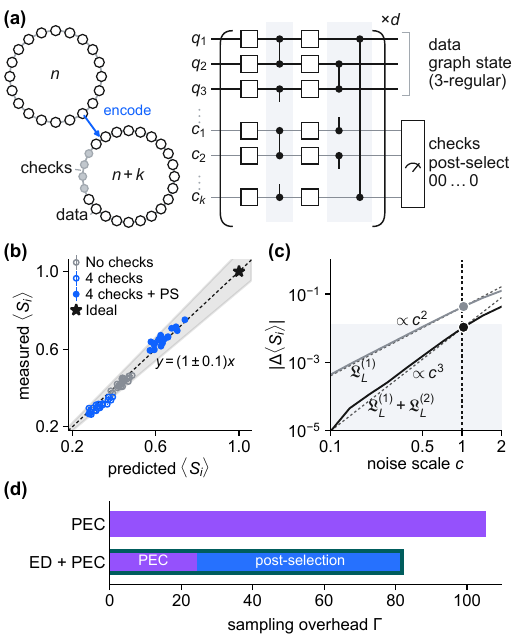}
    \caption{\textbf{Logical noise model validation.}
\textbf{a)} An $n$-qubit graph-state preparation embedded in a ring with $k$ terminal checks and $d$ brickwork repetitions.
\textbf{b)} Predicted and measured values of the 20 graph-state stabilizers $S_i$. The star denotes the ideal stabilizer value $\expval{S_i}=1$. The band denotes $10\%$ relative agreement; error bars are smaller than the markers and are not visible (Appendix~\ref{app:clifford-analysis}).
\textbf{c)} Mean absolute discrepancy between the all-orders Monte Carlo prediction and first- or second-order truncations under rate rescaling $\lambda_\nu\mapsto c\lambda_\nu$. Dashed guides show $c^2$ and $c^3$ scaling; $c=1$ is the physical model. The band extends to twice the mean per-observable standard error of the post-selected data.
\textbf{d)} Inferred overheads, using measured post-selection probability $\alpha=0.2963$. The green outline marks the ED+PEC overhead; purple and blue segments show the PEC contribution and added post-selection cost, respectively.}
    \label{fig:clifford}
\end{figure}

\textit{Protocol validation---}
We first test whether the post-selected logical noise model correctly predicts hardware measurements. On \textit{ibm\_aachen}, the reference circuit prepares a random 3-regular graph state on 20 data qubits using $d=10$ brickwork repetitions, each containing two alternating nearest-neighbor controlled-Z (CZ) layers interleaved with single-qubit Cliffords (Fig.~\ref{fig:clifford}\textbf{a}). The checked circuit prepares the same target state in a 24-qubit ring with $d=12$ repetitions. Its four checks become entangled with the data but ideally return to $\ket{0}^{\otimes 4}$ before measurement. They detect 65\% of the modeled elementary faults in $V$. We measure the stabilizers $S_i=X_i\prod_{j\in\mathcal N(i)}Z_j$, where $\mathcal N(i)$ contains the three neighbors of vertex $i$ of the chosen random graph state. In the ideal noiseless setting $\langle S_i\rangle=1$~(Appendix~\ref{app:clifford-details}).

We learn sparse physical Pauli-Lindblad models from separate calibration circuits~\cite{bergProbabilisticErrorCancellation2023,barron2026observable}, then construct the post-selected logical noise generator through second order (Appendix~\ref{app:clifford-details}). Clifford propagation permits efficient evaluation of the resulting stabilizer predictions~\cite{millerEfficientSimulationClifford2025,hinesSimulatingQuantumError2026}. Figure~\ref{fig:clifford}\textbf{b} compares predictions and measurements without checks, with four checks but no post-selection, and with four checks and post-selection. The mean relative deviations are 3.4\%, 7.3\%, and 3.5\%, respectively, dividing each absolute discrepancy by the magnitude of the corresponding prediction. Post-selection shifts the measured stabilizers toward unity, while the model captures both the additional check-circuit noise and its suppression in the post-selected results.

We test perturbative scaling separately by rescaling the learned gate and check state-preparation and measurement (SPAM) rates, $\lambda_\nu\mapsto c\lambda_\nu$. We compare the truncated models with an all-orders Monte Carlo prediction. The mean absolute truncation error across the 20 stabilizers scales as $c^2$ at first order and $c^3$ at second order (Fig.~\ref{fig:clifford}\textbf{c}). At the physical model strength $c=1$, including $\LL^{(2)}$ reduces this error from $0.04$ to $0.01$, comparable to the experimental uncertainty scale indicated by the grey band~(Appendix~\ref{app:clifford-details}).

The learned models also predict the sampling cost. Without checks, $\Gamma_{\rm PEC}=105.4$. Post-selecting on four checks reduces $\gamma_L^2$ to $24.4$. Including measured post-selection probability $\alpha=0.2963$ gives $\Gamma_{\rm ED+PEC}=82.3$, approximately $22\%$ lower (Fig.~\ref{fig:clifford}\textbf{d}). This inferred reduction includes the checked circuit's greater width and depth. PEC itself is implemented in the next experiment.

\begin{figure}
    \centering
    \includegraphics[width=\columnwidth]{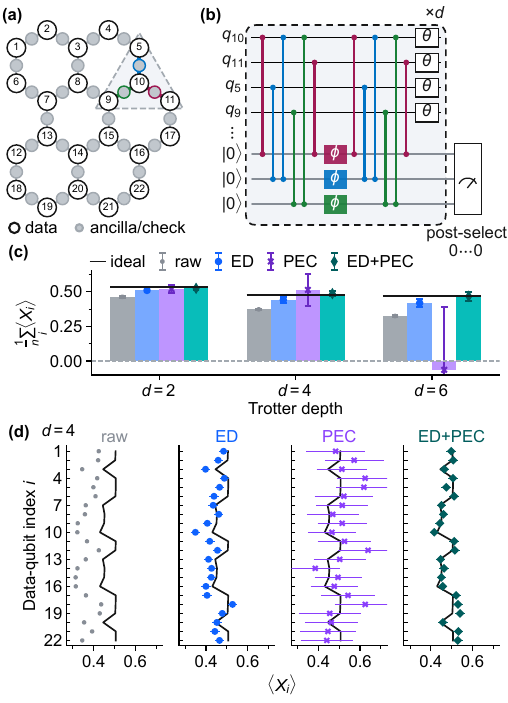}
    \caption{\textbf{Combined error detection and mitigation in a hexagonal Ising simulation}\\ \textbf{a)} Layout of the hexagonal data lattice (numbered qubits) embedded into the heavy-hexagonal lattice.
    \textbf{b)} Example circuit for one Trotter step. Ancilla qubits (lower register) facilitate ZZ-interactions between the connected data qubits. The ancilla boxes implement $R_z(\phi)$ and the data boxes implement $R_x(\theta)$ rotations. 
    \textbf{c)} Mean magnetization $m_x$ at increasing Trotter depths. 
    \textbf{d)} Single-site magnetization $\langle X_i\rangle$ for all 22 data qubits at four Trotter steps. Black lines denote the ideal reference. PEC and ED+PEC use the same number of samples at each depth; raw and ED use smaller budgets. Error bars show one standard error.
    }
    \label{fig:ising_ed_pec}
\end{figure}

\textit{Non-Clifford demonstration---}
We implement Trotterized transverse-field Ising dynamics on a six-plaquette hexagonal lattice using \textit{ibm\_aachen}.  Ancilla qubits implement the $ZZ$ interactions on the hardware's heavy-hex topology and also provide the error-detecting checks (Fig.~\ref{fig:ising_ed_pec}\textbf{a}).  The circuit uses $n=22$ data qubits and $k=27$ check qubits, initialized in $\ket0$ and returned to $\ket0$ after each ideal Trotter step. We measure the checks only at the end and retain the trivial syndrome.

A single Trotter step implements
\begin{equation}
    U_{\rm T} = \prod_{j}\exp\!\left(-i\frac{\theta}{2}X_j\right)
    \prod_{\langle j,k\rangle}\exp\!\left(-i\frac{\phi}{2} Z_j Z_k\right),
    \label{eq:trotter}
\end{equation}
where $\langle j,k\rangle$ denotes neighboring data sites~(Fig.~\ref{fig:ising_ed_pec}\textbf{b}). During the ancilla-assisted $ZZ$ interactions, the propagated checks are $Z$-type at the non-Clifford $R_z(\phi)$ rotations. Between $ZZ$ interactions, they act only on the check register and commute with the data-qubit $R_x(\theta)$ rotations. The circuit thus satisfies the check-commutation condition for deterministic syndromes. Starting from $\ket+^{\otimes n}$, we use $\phi=-\pi/4$ and $\theta=3\pi/8$ for $d=2,4,6$ Trotter steps, with up to 648 CZ gates~(Appendix~\ref{app:ising-details}).

We learn the physical Pauli--Lindblad noise $\Lphys$ from Eq.~\eqref{eq:phys-lind} for each of the three unique $ZZ$ entangling layers following Ref.~\cite{bergProbabilisticErrorCancellation2023}. To reinforce the validity of the Pauli noise model, we detect non-Markovian error processes such as leakage on data and ancilla qubits through a technique introduced in Ref.~\cite{barron2026observable}, on which both the learning circuits and the Ising circuits are post-selected (in addition to any post-selection of the syndrome checks) (Appendix~\ref{app:HC_postselection}). We also rescale expectation values to mitigate state-preparation and measurement (SPAM) errors using the twirled readout error extinction (TREX) technique introduced in Ref.~\cite{van2022model}.

We classify the modeled faults by syndrome and implement $e^{-\LL^{(1)}}$, retaining only undetected generators and sampling the corresponding Pauli branches independently at each layer. The learned models reproduce the measured post-selection rates to about one percentage point, providing a consistency check on the post-selection model (Appendix~\ref{app:syndrome_rate_consistency_check}). First-order inversion leaves contributions from higher-order faults and therefore has a residual bias. We bound the second-order contribution to this bias and find that it remains smaller than the statistical error for all considered observables (Appendix~\ref{app:second_order_bound}).

We compare ED+PEC with error detection alone, PEC applied to the full physical noise model without syndrome post-selection, and an unmitigated baseline. All four methods use the same non-Markovian-error filter and SPAM rescaling. The mean magnetization $m_x=n^{-1}\sum_i\langle X_i\rangle$ summarizes the depth dependence in Fig.~\ref{fig:ising_ed_pec}\textbf{c}. ED+PEC agrees with the ideal mean within the plotted statistical uncertainties at all three depths. At $d=6$, standalone PEC instead deviates strongly from the ideal value and has a much larger uncertainty, illustrating the limited precision attainable with the available samples.

The site-resolved magnetizations at $d=4$ are shown in Fig.~\ref{fig:ising_ed_pec}\textbf{d}. ED reduces the bias relative to the unmitigated result but leaves a residual discrepancy, whereas ED+PEC closely follows the ideal values across the lattice. Standalone PEC again exhibits substantially larger statistical uncertainties and does not converge within the same sampling budget.

Table~\ref{tab:overhead} reports the total sampling overhead $\GammaEDPEC$, including the cost of syndrome post-selection after the common non-Markovian-error filter. ED+PEC reduces the overhead relative to standard PEC by factors of $3.7\times$, $15.9\times$, and $63\times$ at $d=2,4,6$, respectively, demonstrating that the spacetime syndrome classification enables substantial overhead savings and thus enables larger simulations. The improvement factor $\Gamma_{\mathrm{PEC}}/\GammaEDPEC$ generally increases with the circuit volume, while for fixed circuit volumes it becomes larger with the ratio of detected errors. Further experimental details, two-body magnetization results, and model validation are provided in Appendix~\ref{app:ising-details}.

\begin{table}[hbt]
\centering
\caption{
    Sampling overhead for vanilla PEC ($\Gamma_{\mathrm{PEC}}$) and ED+PEC
    ($\GammaEDPEC$) at each Trotter depth on
    \texttt{ibm\_aachen}, conditional on passing the common non-Markovian-error filter.
    The improvement factor is $\Gamma_{\mathrm{PEC}}/\GammaEDPEC$.
}
\label{tab:overhead}
\begin{ruledtabular}
\begin{tabular}{c c c c}
Depth $d$ & \makecell{PEC overhead \\ $\Gamma_{\mathrm{PEC}}$} & 
\makecell{ED+PEC overhead\\ $\GammaEDPEC$} & Improvement \\
\hline
2 &  44.1   &   12.1 &  $3.7\times$ \\
4& 1941    &   122    &  $15.9\times$ \\
6 &  85545   &  1359   &  $63\times$ \\
\end{tabular}
\end{ruledtabular}
\end{table}

\newpage
\textit{Discussion---}
Spacetime PEC characterizes post-selected logical noise and reduces its mitigation cost. While we considered terminal check measurements, a natural extension is to use midcircuit measurements for post-selection. Optimizing their frequency is a promising direction, building on recent work on codesigning error detection and mitigation~\cite{kumarCoDesigningErrorMitigation2026,olearyOptimizingSymmetryInformed2026}. Incorporating these protocols into the present framework requires a consistent treatment of measurement and reset noise, building on advances in midcircuit-measurement noise learning~\cite{zhangGeneralizedCycleBenchmarking2025,hinesPauliNoiseLearning2025}.

Incorporating these operations would also open a route to mitigating residual logical noise after active error correction~\cite{aharonovSyndromeAwareMitigation2025,jeonQuantumErrorCorrection2026}. Under independent stochastic noise, a protocol that corrects all configurations of up to $t$ faults has leading logical errors at order $t+1$ or higher. Extending spacetime PEC to this regime therefore requires efficient higher-order expansions organized by the logical action of faults after decoding. Combining code and decoder structure with logical-noise learning from syndrome data~\cite{zhengEfficientLearningLogical2026,xiaoInSituBenchmarking2026} could make these expansions tractable.

Logical dynamics can be non-Markovian even under Markovian physical noise~\cite{ziyad2025emergent}, although effective Markovian descriptions hold in suitable regimes~\cite{kwiatkowski2025constructing}. The spacetime formulation could capture corrections to these descriptions and correlations from physical noise with memory~\cite{rajmohanCorrelatedCoherentErrors2026}. Finding sparse, learnable representations of these contributions would support mitigation across correlated correction cycles.

\textit{Note added---}
During the preparation of this manuscript, we became aware of related work in Ref.~\cite{yuan2026zeno}, which combines error detection with PEC through Taylor expansion and perturbative inversion of the post-selected channel, with numerical demonstrations on logical Clifford circuits. In contrast, we exponentiate a truncated connected-cumulant expansion to obtain the inverse spacetime noise channel, validate it experimentally, and demonstrate mitigation of non-Clifford Ising dynamics on a superconducting processor.

\begin{acknowledgments}
\textit{Acknowledgements---}We thank Caleb Johnson, Abhinav Kandala, Kristan Temme,
Ewout van den Berg, Ian Hincks, Jordan Hines, Kevin Young, Corey Ostrove, Kenneth Rudinger, and Moein Malekakhlagh for helpful discussions.

The results presented in this work are the subject of a patent application (Application No. 19/365421), filed on 22 Oct. 2025.
\end{acknowledgments}

\appendix
\setcounter{secnumdepth}{2}
\begin{widetext}

\section{Linearized post-selection and PEC normalization}
\label{app:linearized}

The main text assumes Pauli noise in the circuit, which can be obtained through Pauli twirling or randomized compiling~\cite{wallmanNoiseTailoringScalable2016}. These procedures manipulate noise by averaging over randomized circuit implementations. Post-selection requires an additional normalization, whose value can depend on the randomized circuit instance and the input state. Computing an expectation value from the post-selected shots of each randomization and then averaging these values implicitly normalizes each randomization separately. When their post-selection probabilities differ, this procedure generally does not reproduce the post-selected evolution of the averaged circuit, whose logical noise PEC is designed to cancel. We derive a rescaling that restores this agreement. The averaged post-selection probability is state independent in the setting of the main text, yielding a linear post-selected evolution. This appendix derives the rescaling and the corresponding PEC estimator, and illustrates the bias that can arise if the rescaling is omitted.

\subsection{Rescaling post-selected expectations}

Let $r$ label a circuit randomization sampled with probability $p_r$, and let $\mathcal C_r$ denote its noisy evolution. Each randomization implements the same ideal computation. Its unnormalized post-selected map and post-selection probability are
\begin{equation}
    \mathcal B_r(\rho)=\Pi_0\mathcal C_r(\rho)\Pi_0,
    \qquad
    \alpha_r(\rho)=\operatorname{Tr}[\mathcal B_r(\rho)],
    \label{eq:app-accepted-map}
\end{equation}
where $\rho$ includes the initialized check register, as in the main text. For a data observable $O$, extended by the identity on the checks, define $m_r=\operatorname{Tr}[O\mathcal B_r(\rho)]/\alpha_r(\rho)$ whenever $\alpha_r(\rho)>0$, and set $\alpha(\rho)=\sum_r p_r\alpha_r(\rho)$.

A natural analysis procedure is to estimate the observable separately for each randomization using its post-selected shots, then average these estimates with weights $p_r$, which gives $\sum_r p_r m_r$. Computing each conditional mean has already divided that randomization's contribution by its own post-selection probability $\alpha_r(\rho)$. Without post-selection, averaging the per-randomization means reproduces the averaged circuit evolution. With post-selection, however, these separate normalizations generally prevent this identification. Since the PEC model describes the post-selected evolution of the averaged circuit, we must determine how to reconstruct that expectation from the measured conditional means $m_r$. To determine this rescaling, start from the evolution whose logical noise PEC is designed to cancel, $\mathcal C_{\rm phys}=\sum_r p_r\mathcal C_r$, post-selected according to Eq.~\eqref{eq:noisycircuitL}. Linearity of projection and trace gives
\begin{equation}
    \operatorname{Tr}[O\mathcal C_L(\rho)]
    =\frac{\sum_r p_r\operatorname{Tr}[O\mathcal B_r(\rho)]}
    {\sum_r p_r\alpha_r(\rho)}.
\end{equation}
Substituting $\operatorname{Tr}[O\mathcal B_r(\rho)]=\alpha_r(\rho)m_r$ therefore fixes the required weights. For $\alpha(\rho)>0$,
\begin{equation}
    \langle O\rangle_{\rm ps}
    \equiv\operatorname{Tr}[O\mathcal C_L(\rho)]
    =\sum_r p_r\frac{\alpha_r(\rho)}{\alpha(\rho)}m_r.
    \label{eq:linearized-trace}
\end{equation}
Multiplication by $\alpha_r$ restores each randomization's unnormalized contribution, and division by the common $\alpha$ normalizes the average. The relative factor $\alpha_r/\alpha$ is essential here and simply averaging $m_r$ with weights $p_r$ generally describes a different ensemble. When $\alpha_r=0$, post-selection never succeeds for that randomization, so $m_r$ is undefined. Its unnormalized contribution is zero, and it is omitted from sums involving $m_r$.

The rescaling in Eq.~\eqref{eq:linearized-trace} applies to every data observable $O$, so the same construction can be expressed at the level of the post-selected maps. Multiplication by $\alpha_r$ restores each randomization's unnormalized contribution, and averaging these contributions gives
\begin{equation}
    \overline{\mathcal B}(\rho)
    =\sum_r p_r\mathcal B_r(\rho)
    =\Pi_0\mathcal C_{\rm phys}(\rho)\Pi_0.
\end{equation}
The trace of this map is the common post-selection probability $\alpha(\rho)$. Dividing by that trace completes the normalization in Eq.~\eqref{eq:linearized-trace} and gives $\mathcal C_L(\rho)$. This identity holds even when $\alpha$ depends on the input. In the deterministic-syndrome Pauli setting of the main text, $\alpha$ is state independent, so
\begin{equation}
    \overline{\mathcal B}(\rho)=\alpha\,\mathcal C_L(\rho).
\end{equation}
Thus normalization is a constant rescaling of the linear map $\overline{\mathcal B}$, and $\mathcal C_L$ is linear on the logical inputs, whose noise can be mitigated by PEC. 

To implement the rescaling in Eq.~\eqref{eq:linearized-trace} with finite-shot data, we replace the post-selection probabilities and conditional means by their measured estimates. Suppose randomization $r$ is assigned $N_r>0$ attempted shots. Let $N_r^{\rm ps}$ be the number that pass post-selection and $S_r$ the sum of their observable outcomes. Define $\widehat\alpha_r=N_r^{\rm ps}/N_r$, $\widehat m_r=S_r/N_r^{\rm ps}$ for nonzero counts, and $\widehat\alpha=\sum_r p_r\widehat\alpha_r$. Substituting these estimates into Eq.~\eqref{eq:linearized-trace} gives
\begin{equation}
    \widehat{\langle O\rangle}_{\rm ps}
    =\sum_r p_r\frac{\widehat\alpha_r}{\widehat\alpha}\widehat m_r
    =\frac{\sum_r(p_r/N_r)S_r}
    {\sum_r(p_r/N_r)N_r^{\rm ps}}.
    \label{eq:app-allocation-estimator}
\end{equation}
For uniform $p_r$ and equal shot counts, this is the rescaling used in Appendix~\ref{app:clifford-analysis}~\cite{martiel2026sampling}.

\subsection{Amplitude damping example}
\label{app:amplitude-damping}

We illustrate how separate normalization can invalidate the channel description obtained by Pauli twirling. In the following detection circuit, rescaling the post-selected branches by $\alpha_r/\alpha$ before averaging yields a linear logical dephasing channel with an exact PEC inverse. Averaging the normalized branches without rescaling instead yields a nonlinear map, and applying the same inverse leaves a residual bias.

Consider one data qubit in state $\rho_D$ and a check initialized in $\ket0$. Encode with a CNOT from data to check, apply amplitude damping only to the data, and decode with the same CNOT (see Fig.~\ref{fig:scaled_post}\textbf{a}). Post-select the terminal check outcome $0$. All other operations in this example are ideal. With damping parameter $0\leq p<1$, the noise Kraus operators are
\begin{equation}
    E_0=\begin{pmatrix}1&0\\0&\sqrt{1-p}\end{pmatrix},
    \qquad E_1=\sqrt p\ket0\!\bra1.
\end{equation}
After decoding, the no-jump branch has check state $\ket0$ and the jump branch has check state $\ket1$. The post-selected data map is therefore $\rho_D\mapsto E_0\rho_D E_0^\dagger$.

We Pauli-twirl the amplitude-damping channel by uniformly sampling $P\in\{I,X,Y,Z\}$ and applying $P$ immediately before the noise and $P^\dagger$ immediately after it, between encoding and decoding. After decoding and selecting check outcome $0$, the $I/Z$ and $X/Y$ randomizations give two equally weighted unnormalized post-selected maps,
\begin{equation}
    \mathcal B_I^D(\rho_D)=K_I\rho_D K_I^\dagger,
    \quad \mathcal B_X^D(\rho_D)=K_X\rho_D K_X^\dagger,
    \qquad K_I=\begin{pmatrix}1&0\\0&\sqrt{1-p}\end{pmatrix},
    \quad K_X=\begin{pmatrix}\sqrt{1-p}&0\\0&1\end{pmatrix}.
    \label{eq:app-damping-branches}
\end{equation}
Here the superscript $D$ indicates that the post-selected check has been traced out. For $\rho_D=(I+xX+yY+zZ)/2$, with Bloch vector $(x,y,z)$, set
\begin{equation}
    a=1-\frac p2,\qquad b=\frac p2,\qquad
    h=\frac ba=\frac{p}{2-p},\qquad
    f=\frac{\sqrt{1-p}}{a}=\frac{2\sqrt{1-p}}{2-p}.
    \label{eq:app-damping-parameters}
\end{equation}
The two post-selection rates are $\alpha_I=a+bz$ and $\alpha_X=a-bz$. Averaging the unnormalized post-selected maps gives
\begin{equation}
    \frac{\mathcal B_I^D(\rho_D)+\mathcal B_X^D(\rho_D)}2
    =a\,\mathcal D_f(\rho_D),
    \qquad
    \mathcal D_f(\rho_D)=\frac{I+fxX+fyY+zZ}{2}.
    \label{eq:app-pooled-damping}
\end{equation}
where $\mathcal D_f$ is the logical dephasing channel. In contrast, averaging the two separately normalized states gives a map $\mathcal F_p$ with Bloch action
\begin{equation}
    \mathcal F_p:\ (x,y,z)\longmapsto
    \left(\frac{fx}{1-h^2z^2},\frac{fy}{1-h^2z^2},
    \frac{(1-h^2)z}{1-h^2z^2}\right).
    \label{eq:app-separate-damping}
\end{equation}
Its input-dependent denominator illustrates the issue with the nonlinearity of post-selection. For diagonal inputs, both $Z$ eigenstates are fixed, but their mixtures are generally displaced from the identity line in Fig.~\ref{fig:scaled_post}\textbf{b}.

The inverse of the logical channel has the quasiprobability decomposition
\begin{equation}
    \mathcal D_f^{-1}(\cdot)
    =q_I(\cdot)+q_Z Z(\cdot)Z,
    \qquad q_I=\frac{1+f^{-1}}2,\quad q_Z=\frac{1-f^{-1}}2.
    \label{eq:app-damping-inverse}
\end{equation}
After normalization by the common post-selection probability $\alpha=a$, applying $\mathcal D_f^{-1}$ recovers $\rho_D$ exactly. Applying the same modeled inverse to $\mathcal F_p$, however, leaves
\begin{equation}
    \Delta x=\frac{x h^2z^2}{1-h^2z^2},\qquad
    \Delta y=\frac{y h^2z^2}{1-h^2z^2},\qquad
    \Delta z=-\frac{z h^2(1-z^2)}{1-h^2z^2}.
    \label{eq:app-damping-bias}
\end{equation}
These biases are generally $O(p^2)$ at small $p$ and arise from using the inverse of the linear logical channel on a different, nonlinear ensemble (see Fig.~\ref{fig:scaled_post}\textbf{c}). Relative-probability rescaling avoids this mismatch. In this example the $I/Z$ corrections can be applied after decoding and preserve the post-selection probability, giving
\begin{equation}
    \gamma=|q_I|+|q_Z|=f^{-1},\qquad
    \Gamma_{\rm ED+PEC}=\frac{\gamma^2}{\alpha}
    =\frac{2-p}{2(1-p)}.
    \label{eq:app-damping-overhead}
\end{equation}

\begin{figure}[t]
    \centering
    \includegraphics[width=0.94\linewidth]{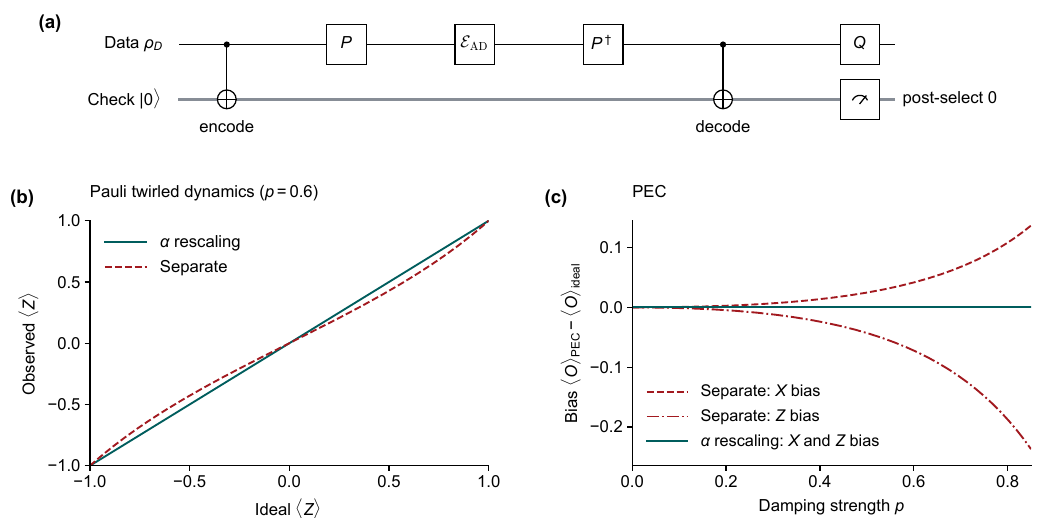}
    \caption{\textbf{Rescaling by post-selection probabilities enables modeled PEC.} \textbf{a)} Ideal error detection circuit with amplitude damping $\mathcal E_{\rm AD}$ on the data. A uniformly sampled Pauli $P\in\{I,X,Y,Z\}$ and its inverse $P^\dagger$ surround the noise between encoding and decoding, implementing Pauli twirling. The two procedures act on the same unnormalized post-selected maps. The check outcome $0$ selects the no-jump branch. For panel \textbf{b}, $Q=I$, and for panel \textbf{c}, the PEC correction $Q\in\{I,Z\}$ is sampled with probability $|q_Q|/\gamma$ and its measurement outcome weighted by $\gamma\operatorname{sgn}(q_Q)$, using Eq.~\eqref{eq:app-damping-inverse}. \textbf{b)} Exact output $\langle Z\rangle$ for diagonal inputs $\rho_D=(I+zZ)/2$ at $p=0.6$. Rescaling with post-selection probability preserves the identity line, while averaging separately normalized branches deviates from the expected behavior. \textbf{c)} Exact residual biases after applying $\mathcal D_f^{-1}$, for the pure input with Bloch vector $(\sqrt3/2,0,1/2)$. Separate normalization leaves both X and Z biases, whereas rescaling gives zero bias for both. Curves are analytic predictions, with no shot noise.}
    \label{fig:scaled_post}
\end{figure}

\subsection{Probabilistic error cancellation and sampling overhead}
\label{app:pec-estimator}

PEC implements the inverse noise model on average by sampling Pauli operations and multiplying the measurement outcomes by the sampled sign and the sampling overhead factor~\cite{bergProbabilisticErrorCancellation2023}. The inverse of the logical noise model constructed in the main text defines a quasiprobability distribution over Pauli operations,
\begin{equation}
    e^{-\LL}(\cdot)
    =\sum_c q_c\,\mathbb Q_c(\cdot)\mathbb Q_c^\dagger,
    \qquad \sum_c q_c=1,
    \qquad \gamma_L=\sum_c|q_c|.
\end{equation}
Here $c$ labels a complete choice of Pauli operations sampled from the inverse factors, and $\mathbb Q_c$ records their circuit locations in the auxiliary space. The coefficients $q_c$ can be negative. We sample $c$ with probability $|q_c|/\gamma_L$, insert its Pauli operations at the specified locations, and multiply the measurement outcome by $\gamma_L\operatorname{sgn}(q_c)$. Operations at different locations may need to be sampled jointly, as described in the main text.

Combining PEC with error detection uses the same rescaling as Eq.~\eqref{eq:linearized-trace}. Let $\tau$ label a twirling randomization with probability $p_\tau$, and $m_{\tau,c}$ its normalized post-selected mean with correction $c$. The desired signed expectation is
\begin{equation}
    m_{\rm PEC}=\sum_{\tau,c}p_\tau q_c
    \frac{\alpha_{\tau,c}}{\alpha}m_{\tau,c}.
    \label{eq:app-rescaled-pec}
\end{equation}
Each factor $\alpha_{\tau,c}$ restores an unnormalized contribution before the signed average; the common $\alpha$ then normalizes the result. Zero-probability branches contribute zero.

To identify the common normalization, we assume that the correction coefficients $q_c$ are independent of $\tau$ and that inserting the corrections does not change the noise model. In the  Pauli-twirled model, every sampled spacetime correction
has zero total syndrome and therefore preserves the syndrome of each
physical fault trajectory. Consequently,
$\sum_\tau p_\tau\alpha_{\tau,c}=\alpha$ for every sampled branch $c$.
Using $\sum_c q_c=1$ and $\sum_c|q_c|=\gamma_L$, it follows that
\begin{equation}
    \sum_{\tau,c}p_\tau q_c\alpha_{\tau,c}
    =\sum_{\tau,c}p_\tau\frac{|q_c|}{\gamma_L}\alpha_{\tau,c}
    =\alpha.
    \label{eq:app-unsigned-normalizer}
\end{equation}
The normalization required by the inverse model is therefore the post-selection probability of the sampled circuits. For $T$ shots, let $A_t\in\{0,1\}$ indicate successful post-selection, $\sigma_t=\operatorname{sgn}(q_{c_t})$, and $o_t\in[-1,1]$ be the observable outcome, with $A_to_t=0$ for rejected shots. Replacing $\alpha$ by $\widehat\alpha=T^{-1}\sum_t A_t$ gives
\begin{equation}
    \widehat m_{\rm PEC}
    =\frac{1}{T\widehat\alpha}\sum_t\gamma_L\sigma_t A_t o_t
    =\gamma_L\frac{\sum_t\sigma_t A_t o_t}{\sum_t A_t}.
    \label{eq:app-pooled-pec}
\end{equation}
This count formula implements the rescaling in Eq.~\eqref{eq:app-rescaled-pec}; its denominator must be nonzero. An exact inverse removes the modeled logical noise bias; using a truncated logical generator can leave the residual bias discussed in the main text.

The factor $\gamma_L$ also determines the sampling cost. For $T$ independent, identically distributed shots, let $W=\gamma_L\sigma o$ denote the weighted outcome of a post-selected shot. When the twirl and PEC correction are sampled independently for each shot, the rescaled estimator in Eq.~\eqref{eq:app-pooled-pec} is unbiased conditional on at least one post-selected shot. More general shot allocations can introduce finite-sample ratio bias. For a sufficiently large expected number of post-selected shots, $T\alpha$, its variance conditional on a nonzero post-selected count is approximately
\begin{equation}
    \operatorname{Var}(\widehat m_{\rm PEC})
    \approx\frac{\operatorname{Var}(W\mid A=1)}{T\alpha}
    \leq\frac{\gamma_L^2}{T\alpha}.
\end{equation}
Consequently, the sampling overhead used in the main text is $\Gamma_{\rm ED+PEC}=\gamma_L^2/\alpha$, where PEC contributes $\gamma_L^2$, and post-selection contributes $1/\alpha$. For the factorized inverse of a logical generator with coefficients $\kappa_j$, each inverse factor contributes $\exp[2\max(\kappa_j,0)]$, giving $\gamma_L=\exp[2\sum_j\max(\kappa_j,0)]$.

\section{Perturbative construction of the post-selected logical generator}
\label{app:higher-order}

We construct the generator of the post-selected logical channel by exponentiating the physical generator, post-selecting zero-syndrome configurations, normalizing by their total weight, and taking the logarithm. Individually undetected faults factor out exactly. The remaining logarithmic expansion begins at second order in the detected-fault rates.

\subsection{Syndrome rules and zero-syndrome configurations}
\label{app:syndrome-rule}
\label{app:spacetime-pec}

We use the setting introduced in the main text, with circuit $U=U_M\cdots U_1$ on $N=n+k$ qubits, where faults occur after their labeled layer. Let $c_j$, $j=1,\dots,k$, denote the check qubits. The backward image of the terminal check $Z_{c_j}$ at a fault location after layer $\ell$ is
\begin{equation}
G_j^{(\ell)}
=U_{\ell+1}^\dagger\cdots U_M^\dagger Z_{c_j}
U_M\cdots U_{\ell+1},
\qquad
\mathbb G_{\bm b_j}=\bigotimes_{\ell=1}^M G_j^{(\ell)}.
\label{eq:app-backward-check}
\end{equation}
For a Clifford circuit, these images are Paulis. Non-Clifford gates are also allowed provided that every back-propagated check commutes with each non-Clifford gate it encounters, leaving the check unchanged under conjugation.

For an elementary fault $\nu=(a,\ell)\in V$, let $\Pst_\nu$ act as $P_a$ at layer $\ell$ and as identity elsewhere, i.e., $
    \Pst_\nu
    =
    I^{\otimes(\ell-1)}
    \otimes P_a
    \otimes I^{\otimes(M-\ell)}$. Its syndrome bit is
\begin{equation}
s_{\nu,j}=
\begin{cases}
0,&[\mathbb G_{\bm b_j},\Pst_\nu]=0,\\
1,&\{\mathbb G_{\bm b_j},\Pst_\nu\}=0.
\end{cases}
\label{eq:syndrome-commutation}
\end{equation}
Since the ideal output has check eigenvalue $+1$ for every allowed data input, the syndrome is deterministic and independent of the data state.

Each elementary generator factor contributes either its identity or Pauli branch, as derived below. A fault configuration is a subset $A\subseteq V$ specifying the labels for which the Pauli branch is selected. Multiplying the selected branches at each layer gives the corresponding trajectory $\bm a$, whose auxiliary Pauli trajectory and syndrome are
\begin{equation}
\mathbb Q_A=\prod_{\nu\in A}\Pst_\nu,
\qquad
\bm s(A)=\bigoplus_{\nu\in A}\bm s_\nu,
\label{eq:app-configuration}
\end{equation}
where irrelevant Pauli phases are omitted and $\oplus$ denotes bitwise addition modulo two. The syndrome sum follows because the individual commutation signs multiply under the product $\mathbb Q_A$. Configurations with $\bm s(A)=\bm 0$ survive post-selection, including collections of individually detected faults whose syndromes cancel. The subset $A$ is distinct from the resulting trajectory $\faulttraj$. Different subsets can yield the same Pauli at every layer, so their weights must be summed when forming the trajectory probabilities used in the main text.

\subsection{Complexity analysis}

We first consider Clifford circuits. The syndrome of an elementary Pauli fault can be obtained by propagating the check observables backward and computing their commutation relation with the fault in the spacetime picture, as in Eq.~\eqref{eq:syndrome-commutation}.

Equivalently, one could propagate each Pauli fault forward to the end of the circuit and compute its commutation relation with each terminal check observable $Z_{c_j}$.
These two approaches have different computational costs. Our approach, inspired by the construction of Ref.~\cite{delfosse2023simulationnoisycliffordcircuits}, relies on computing $k$ check trajectories, each costing $O(MN)$ operations. Commutation with a local fault can then be computed in constant time for a given fault and check trajectory. Overall, collecting the syndromes of the $|V|$ elementary fault labels has complexity $O(k(MN+|V|))$.
Computing the same syndromes by propagating faults forward interchanges the roles of the $|V|$ faults and the $k$ checks, giving complexity $O(|V|(MN+k))$. In our setting, $|V|\gg k$, which motivates propagating the checks backward.

\subsection{Presence of non-Clifford Pauli rotations}
\label{app:nonclifford}

We now extend the syndrome construction to circuits containing non-Clifford Pauli rotations that commute with the back-propagated check observables. For clarity, consider a single non-Clifford Pauli rotation of angle $\theta$ about axis $Q$ and a Pauli fault $P$ acting before this rotation, with $\{P,Q\}=0$. We can propagate $P$ forward past the rotation, giving a new operator $\tilde P$:
\begin{align*}
    \tilde{P}&=R_Q(\theta) \cdot P \cdot R_Q(-\theta) \\
    &= \left(\cos({\theta/2}) I -i \sin(\theta/2)Q  \right)\cdot P\cdot \left(\cos({\theta/2}) I + i \sin(\theta/2)Q  \right)\\
    &=\left(\cos^2(\theta/2) - \sin^2(\theta/2)\right) P + 2i\sin(\theta/2)\cos(\theta/2)PQ\\
    &=\cos\theta\,P+i\sin\theta\,PQ.
\end{align*}
The syndrome of this new operator, placed after the rotation, is the same as that of the original fault $P$. Let $G=G_j^{(\ell)}$ denote the physical back-propagated image of check $j$ immediately after the rotation. By assumption, $[G,Q]=0$, so the rotation leaves this check image unchanged. Thus,
\begin{align*}
    [G,\tilde P] &= \cos\theta\,[G,P]+i\sin\theta\,[G,PQ]\\
                 &= \cos\theta\,[G,P]+i\sin\theta\,[G,P]Q\\
                 &= [G,P](\cos\theta\,I+i\sin\theta\,Q)\\
                 &= [G,P]e^{i\theta Q}.
\end{align*}

Since $e^{i\theta Q}$ is unitary, $[G,P]=0$ if and only if $[G,\tilde P]=0$. Similarly, $\{G,\tilde P\}=\{G,P\}e^{i\theta Q}$, so anticommutation is also preserved. Both syndrome values are therefore unchanged. If $[P,Q]=0$, the rotation leaves $P$ itself unchanged, and the conclusion follows directly.

\subsection{Fault weights and the post-selected channel}

Define conjugation maps on auxiliary operators by
\begin{equation}
    \mathfrak P_\nu(\cdot)=\Pst_\nu(\cdot)\Pst_\nu,
    \qquad \mathfrak P_A=\prod_{\nu\in A}\mathfrak P_\nu,
    \qquad \mathfrak I(\cdot)=(\cdot).
\end{equation}
Although the Pauli operators may anticommute, their conjugation maps commute because the phases cancel. Also $\mathfrak P_\nu^2=\mathfrak I$. The physical generator of Eq.~\eqref{eq:phys-lind} therefore exponentiates as
\begin{align}
    e^{\Lphys}
    &=\prod_{\nu\in V}e^{\lambda_\nu(\mathfrak P_\nu-\mathfrak I)},\\
    e^{\lambda_\nu(\mathfrak P_\nu-\mathfrak I)}
    &=e^{-\lambda_\nu}
      (\cosh\lambda_\nu\,\mathfrak I+\sinh\lambda_\nu\,\mathfrak P_\nu)
      =(1-p_\nu)\mathfrak I+p_\nu\mathfrak P_\nu,
    \qquad p_\nu=\frac{1-e^{-2\lambda_\nu}}2.
    \label{eq:app-binary-factor}
\end{align}
Thus
\begin{equation}
    e^{\Lphys}=\sum_{A\subseteq V}w(A)\mathfrak P_A,
    \qquad
    w(A)=\prod_{\nu\in A}p_\nu\prod_{\nu\notin A}(1-p_\nu).
    \label{eq:app-configuration-weights}
\end{equation}
For $\lambda_\nu\geq0$, $p_\nu$ is a Bernoulli probability and $w(A)$ is the probability of configuration $A$. For signed rates, the same identities hold algebraically, but the elementary weights need not be nonnegative~\cite{seif2026single}.

The unnormalized post-selected auxiliary map, its post-selection probability, and its normalized channel are
\begin{equation}
    \mathfrak B_{\bm0}
    =\sum_{A:\,\bm s(A)=\bm0}w(A)\mathfrak P_A,
    \qquad
    \alpha=\sum_{A:\,\bm s(A)=\bm0}w(A),
    \qquad
    e^{\LL}=\frac{\mathfrak B_{\bm0}}{\alpha}.
    \label{eq:app-normalized-auxiliary}
\end{equation}
Here $\alpha$ is state independent because each configuration has a fixed syndrome. Summing configuration weights $w(A)$ over configurations that produce the same trajectory gives $\probphys(\bm a)$. Restricting to post-selected trajectories and dividing by $\alpha$ gives $\probL(\bm a)$, which determines the physical post-selected evolution $\mathcal C_L$ defined in Eq.~\eqref{eq:noisycircuitL}. The post-selected auxiliary channel $e^{\LL}$ encodes these probabilities and does not act on the physical state.

\subsection{Exact factorization of the post-selected channel}

As in the main text, separate the elementary labels into individually undetected faults $\Vundet$ and detected faults $\Vdet$. Write each configuration as $A=D\cup U$, with $D\subseteq\Vdet$ and $U\subseteq\Vundet$. Since undetected faults contribute zero syndrome, $\bm s(A)=\bm s(D)$, so post-selection constrains only $D$. The auxiliary channel representing physical noise factors as
\begin{equation}
    e^{\Lphys}
    =e^{\Lundet}\prod_{\nu\in\Vdet}
      \bigl[(1-p_\nu)\mathfrak I+p_\nu\mathfrak P_\nu\bigr],
    \qquad
    \Lundet=\sum_{\nu\in\Vundet}\lambda_\nu(\mathfrak P_\nu-\mathfrak I).
\end{equation}
We therefore retain $e^{\Lundet}$ unchanged and apply the zero-syndrome selection only to the detected-fault product.

To expand the detected-fault product, write each factor as
\begin{equation}
    (1-p_\nu)\mathfrak I+p_\nu\mathfrak P_\nu
    =(1-p_\nu)(\mathfrak I+t_\nu\mathfrak P_\nu),
    \qquad t_\nu=\frac{p_\nu}{1-p_\nu}=\tanh\lambda_\nu.
\end{equation}
Writing $t_D=\prod_{\nu\in D}t_\nu$, selecting the zero-syndrome terms gives the unnormalized post-selected map
\begin{equation}
    \mathfrak B_{\bm0}
    =e^{\Lundet}\left[\prod_{\nu\in\Vdet}(1-p_\nu)\right]
      \mathfrak B_{\rm d},
    \qquad
    \mathfrak B_{\rm d}
    =\sum_{\substack{D\subseteq\Vdet\\\bm s(D)=\bm0}}
       t_D\mathfrak P_D,
    \label{eq:app-exact-factorization}
\end{equation}
with $t_\varnothing=1$ and $\mathfrak P_\varnothing=\mathfrak I$.

To obtain the post-selection probability, we sum the weights over the undetected configurations. For fixed $D$, this gives
\begin{equation}
\begin{aligned}
    \sum_{U\subseteq\Vundet}w(D\cup U)
    &=\prod_{\nu\in D}p_\nu
      \prod_{\nu\in\Vdet\setminus D}(1-p_\nu)
      \prod_{\nu\in\Vundet}[(1-p_\nu)+p_\nu]\\
    &=\left[\prod_{\nu\in\Vdet}(1-p_\nu)\right]t_D.
\end{aligned}
\end{equation}
For each undetected label, the sum includes both possibilities. The fault is absent with weight $1-p_\nu$, or present with weight $p_\nu$. Their sum is one, so summing over all undetected configurations leaves only the detected-fault weight. To obtain the second line, write $p_\nu=(1-p_\nu)t_\nu$ for each $\nu\in D$. Every detected label then contributes a factor $1-p_\nu$, while the labels in $D$ additionally contribute $t_D$. Finally, summing over all $D$ with zero syndrome gives the post-selection probability.
\begin{equation}
    \alpha=\left[\prod_{\nu\in\Vdet}(1-p_\nu)\right]z_{\rm d},
    \qquad
    z_{\rm d}=\sum_{\substack{D\subseteq\Vdet\\\bm s(D)=\bm0}}t_D.
    \label{eq:app-reduced-sums}
\end{equation}
Dividing the post-selected map by $\alpha$ cancels the common product of prefactors and gives the normalized post-selected auxiliary channel
\begin{equation}
    e^{\LL}=e^{\Lundet}\frac{\mathfrak B_{\rm d}}{z_{\rm d}}.
    \label{eq:app-normalized-factorization}
\end{equation}

We obtain the post-selected logical generator by taking the logarithm and expanding about zero rates. Since the auxiliary maps commute, their logarithms add.
\begin{equation}
    \LL=\Lundet+\Ldet,\qquad
    \Ldet=\log\mathfrak B_{\rm d}-(\log z_{\rm d})\mathfrak I.
    \label{eq:app-normalized-logarithm}
\end{equation}
The scalar logarithm accounts for normalization and ensures that the generator has zero trace action.

Individually undetected faults therefore retain their original generator $\Lundet=\LL^{(1)}$ exactly, as in Eq.~\eqref{eq:first_order_generator}, with no mixed undetected/detected terms in the generator. The contribution $\Ldet$ comes solely from zero-syndrome combinations of detected faults. Its expansion begins at second order because no single detected fault survives post-selection.

\subsection{Generator expansion through fourth order}

We now expand the detected-fault contribution $\Ldet$ to recover the second-order generator given in the main text and derive the third- and fourth-order corrections. Recall that we organize the logical generator by total degree in the physical rates,
\begin{equation}
    \LL=\LL^{(1)}+\sum_{j\geq2}\LL^{(j)},
    \qquad \LL^{(1)}=\Lundet,
\end{equation}
We obtain these terms by expanding $\mathfrak B_{\rm d}$ and $z_{\rm d}$ in Eq.~\eqref{eq:app-normalized-logarithm}. Their superscripts $(j)$ likewise denote contributions of degree $j$. Since a single detected fault does not survive post-selection, the first corrections to $\mathfrak B_{\rm d}$ and $z_{\rm d}$ are second order
\begin{equation}
    \mathfrak B_{\rm d}
    =\mathfrak I+\mathfrak B_{\rm d}^{(2)}
      +\mathfrak B_{\rm d}^{(3)}+\mathfrak B_{\rm d}^{(4)}+O(\lambda_d^5),
    \qquad
    z_{\rm d}=1+z_{\rm d}^{(2)}+z_{\rm d}^{(3)}+z_{\rm d}^{(4)}+O(\lambda_d^5),
\end{equation}
where $\lambda_d$ denotes the characteristic magnitude of the detected-fault rates as in the main text. Expanding the two logarithms in Eq.~\eqref{eq:app-normalized-logarithm} gives the second-, third-, and fourth-order contributions to the generator
\begin{align}
    \LL^{(2)}&=\mathfrak B_{\rm d}^{(2)}-z_{\rm d}^{(2)}\mathfrak I,\label{eq:app-log-order2}\\
    \LL^{(3)}&=\mathfrak B_{\rm d}^{(3)}-z_{\rm d}^{(3)}\mathfrak I,\label{eq:app-log-order3}\\
    \LL^{(4)}&=\mathfrak B_{\rm d}^{(4)}
       -\frac12(\mathfrak B_{\rm d}^{(2)})^2
       -\left[z_{\rm d}^{(4)}-\frac12(z_{\rm d}^{(2)})^2\right]\mathfrak I.
    \label{eq:app-log-order4}
\end{align}
Exponentiating the second-order generator already produces fourth-order terms through $\tfrac12(\LL^{(2)})^2$. Taking the logarithm subtracts these products, leaving the connected fourth-order contribution $\LL^{(4)}$.

To find the coefficients explicitly, let $\Omega_r$ contain the zero-syndrome configurations with $r$ distinct detected labels, and define their rate products by
\begin{equation}
    \Omega_r=\{D\subseteq\Vdet:\ |D|=r,\ \bm s(D)=\bm0\},
    \qquad \lambda_D=\prod_{\nu\in D}\lambda_\nu.
\end{equation}
Since $t_\nu=\lambda_\nu-\lambda_\nu^3/3+O(\lambda_\nu^5)$,
\begin{equation}
\begin{aligned}
    \mathfrak B_{\rm d}^{(2)}
      &=\sum_{D\in\Omega_2}\lambda_D\mathfrak P_D,
    & z_{\rm d}^{(2)}&=\sum_{D\in\Omega_2}\lambda_D,\\
    \mathfrak B_{\rm d}^{(3)}
      &=\sum_{D\in\Omega_3}\lambda_D\mathfrak P_D,
    & z_{\rm d}^{(3)}&=\sum_{D\in\Omega_3}\lambda_D.
\end{aligned}
\end{equation}
Substituting these expressions into Eqs.~\eqref{eq:app-log-order2}~and~\eqref{eq:app-log-order3} gives the second- and third-order generators
\begin{align}
    \LL^{(2)}
    &=\sum_{\{\nu,\mu\}\in\Omega_2}\lambda_\nu\lambda_\mu
      (\mathfrak P_\mu\mathfrak P_\nu-\mathfrak I),
    \label{eq:app-second-order-revised}\\
    \LL^{(3)}
    &=\sum_{D\in\Omega_3}\lambda_D(\mathfrak P_D-\mathfrak I).
    \label{eq:app-third-order-revised}
\end{align}
Each unordered pair is counted once. Using $\mathfrak P_\mu\mathfrak P_\nu(\cdot) =\Pst_\mu\Pst_\nu(\cdot)\Pst_\nu\Pst_\mu$, the second-order expression is equivalent to the ordered sum with prefactor $1/2$ in Eq.~\eqref{eq:logical-generator-second-order-main} of the main text.

At fourth order, the weights of zero-syndrome configurations receive contributions from two sources. For a configuration of four detected faults, taking the linear term from each $t_\nu$ gives $\lambda_D$. For a zero-syndrome pair, taking the cubic term from one weight and the linear term from the other gives
\begin{equation}
    t_\nu t_\mu=\lambda_\nu\lambda_\mu
      -\frac13(\lambda_\nu^3\lambda_\mu+\lambda_\nu\lambda_\mu^3)
      +O(\lambda_d^6).
\end{equation}
A zero-syndrome triple has leading degree three, and replacing one linear factor with a cubic factor raises its degree to five. It therefore makes no fourth-order contribution. The fourth-order map and normalization terms are
\begin{equation}
\begin{aligned}
    \mathfrak B_{\rm d}^{(4)}
    &=\sum_{D\in\Omega_4}\lambda_D\mathfrak P_D
    -\frac13\sum_{\{\nu,\mu\}\in\Omega_2}
       (\lambda_\nu^3\lambda_\mu+\lambda_\nu\lambda_\mu^3)
       \mathfrak P_\mu\mathfrak P_\nu,\\
    z_{\rm d}^{(4)}
    &=\sum_{D\in\Omega_4}\lambda_D
    -\frac13\sum_{\{\nu,\mu\}\in\Omega_2}
       (\lambda_\nu^3\lambda_\mu+\lambda_\nu\lambda_\mu^3).
\end{aligned}
\end{equation}

To obtain the generator, we also include the quadratic terms of the two logarithms. Because the corrections to $\mathfrak B_{\rm d}$ and $z_{\rm d}$ begin at second order, only the squares of their second-order terms contribute at fourth order. Equation~\eqref{eq:app-log-order4} therefore gives
\begin{equation}
    \LL^{(4)}
    =\mathfrak B_{\rm d}^{(4)}-z_{\rm d}^{(4)}\mathfrak I
     -\frac12\left[
       (\mathfrak B_{\rm d}^{(2)})^2
       -(z_{\rm d}^{(2)})^2\mathfrak I
     \right].
\end{equation}
Squaring the second-order sums produces a double sum over zero-syndrome pairs. Substituting their weights and the fourth-order expressions above yields
\begin{equation}
\begin{aligned}
    \LL^{(4)}={}&\sum_{D\in\Omega_4}\lambda_D(\mathfrak P_D-\mathfrak I)\\
    &-\frac13\sum_{\{\nu,\mu\}\in\Omega_2}
       (\lambda_\nu^3\lambda_\mu+\lambda_\nu\lambda_\mu^3)
       (\mathfrak P_\mu\mathfrak P_\nu-\mathfrak I)\\
    &-\frac12\sum_{D,D'\in\Omega_2}\lambda_D\lambda_{D'}
       (\mathfrak P_D\mathfrak P_{D'}-\mathfrak I).
\end{aligned}
\label{eq:app-fourth-order-revised}
\end{equation}
The three lines contain the four-fault contributions, the corrections to weights of zero-syndrome pairs, and the subtraction of products of second-order contributions, respectively. The last sum runs over ordered choices of zero-syndrome pairs, with the factor $1/2$ supplied by the logarithm. The pairs may overlap or coincide. For $D=D'$ the map product is $\mathfrak I$ and the summand vanishes; for overlapping pairs, repeated labels cancel in the map product but remain in the rate product. The full expansion through this order is $\LL=\Lundet+\LL^{(2)}+\LL^{(3)}+\LL^{(4)}+O(\lambda_d^5)$.

\section{Details on the Clifford experiment}
\label{app:clifford-details}

\subsection{Target state, circuits, and device layout}
\label{app:clifford-layout}

The Clifford experiment in Fig.~\ref{fig:clifford} tests the learned post-selected logical noise model on \textit{ibm\_aachen}. Both circuits prepare the same 20-qubit, 3-regular graph state. Its canonical stabilizers are $S_i=X_i\prod_{j\in\mathcal N(i)}Z_j$, with logical data indices $i=0,\ldots,19$ and neighborhoods specified in Table~\ref{tab:clifford-graph}. All 20 ideal expectation values are $+1$. The reference circuit uses $N=20$ qubits and $d=10$ brickwork repetitions; the checked circuit uses $N=24$ qubits and $d=12$ repetitions, with qubits $20,\ldots,23$ serving as checks. The checks are initialized in $\ket{0}^{\otimes 4}$ and return to that state in the ideal preparation.

Each repetition consists of two alternating layers of nearest-neighbor CZ gates on a ring, interleaved with single-qubit Clifford gates. The gate pairs in these layers are $\{(0,1),(2,3),\ldots,(N-2,N-1)\}$ and $\{(1,2),(3,4),\ldots,(N-1,0)\}$. There are $N$ CZ gates per repetition, giving 200 CZ gates in 20 entangling layers for the reference and 288 CZ gates in 24 entangling layers for the checked preparation.

Fig.~\ref{fig:clifford-layout} shows the layout and qubit mapping in the experiment. The two preparations have different mappings and their common graph state is defined by the logical data indices.

\begin{table}[b]
\centering
\caption{Neighborhoods of the target 3-regular graph. These define the 20
canonical stabilizers measured in the Clifford experiment.}
\label{tab:clifford-graph}
\begin{ruledtabular}
\begin{tabular}{c c @ {\qquad\qquad} c c}
$i$ & $\mathcal N(i)$ & $i$ & $\mathcal N(i)$\\
\hline\\
0 & $\{1,9,18\}$ & 10 & $\{2,15,16\}$ \\
1 & $\{0,7,14\}$ & 11 & $\{5,12,14\}$ \\
2 & $\{3,4,10\}$ & 12 & $\{3,9,11\}$ \\
3 & $\{2,8,12\}$ & 13 & $\{7,17,18\}$ \\
4 & $\{2,8,15\}$ & 14 & $\{1,8,11\}$ \\
5 & $\{11,16,19\}$ & 15 & $\{4,10,16\}$ \\
6 & $\{17,18,19\}$ & 16 & $\{5,10,15\}$ \\
7 & $\{1,9,13\}$ & 17 & $\{6,13,19\}$ \\
8 & $\{3,4,14\}$ & 18 & $\{0,6,13\}$ \\
9 & $\{0,7,12\}$ & 19 & $\{5,6,17\}$ \\
\end{tabular}
\end{ruledtabular}
\end{table}

\begin{figure}
    \centering
    \includegraphics[width=0.93\linewidth]{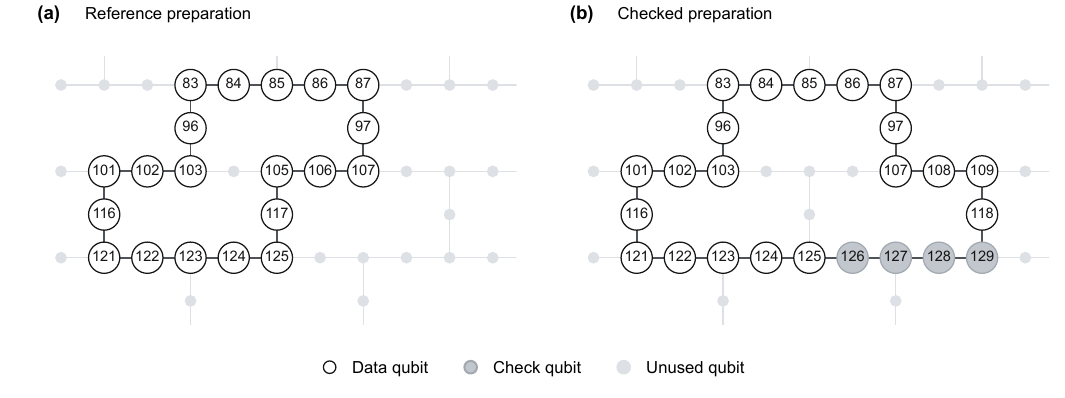}
    \caption{\textbf{Qubit mappings for the Clifford experiment.}
    \textbf{a)} Reference preparation with 20 data qubits.
    \textbf{b)} Checked preparation with 20 data and four check qubits.
    Numbers inside the nodes are physical \textit{ibm\_aachen} qubit
    identifiers.
    White circles denote data, gray circles denote checks, and pale gray
    nodes and edges show unused parts of the device. Dark edges show the
    connections used by the preparation circuit. The regular heavy-hex
    embedding preserves device connectivity; distances are schematic.}
    \label{fig:clifford-layout}
\end{figure}

\subsection{Measurement settings, randomizations, and shot budget}
\label{app:clifford-shots}

The canonical stabilizers are partitioned into three qubit-wise commuting groups, so every group can be measured with a single tensor-product basis:
\begin{equation}
\begin{aligned}
    \mathcal G_1&=\{0,2,5,6,7,8,12,15\},\\
    \mathcal G_2&=\{1,3,4,9,10,11,13,19\},\\
    \mathcal G_3&=\{14,16,17,18\}.
\end{aligned}
\label{eq:clifford-groups}
\end{equation}
A label $i\in\mathcal G_g$ denotes $S_i$. For each of the three settings, we use 500 circuit randomizations with 100 attempted shots per randomization. Randomization includes Pauli twirling of the entangling layers and measurement randomization, with the terminal Pauli frame corrected in classical processing.

The checked data are analyzed both with and without post-selection; these two plotted conditions use the same acquired shots. We post-select on all four corrected check bits being zero. The three measurement settings retain 14576, 15238, and 14631 shots, respectively, giving $\alpha=0.2963$.

Table~\ref{tab:clifford-budget} lists the measurement budget, including the independent noise-characterization circuits. Characterization circuits also use the same 500 randomizations and 100 shots per randomization.

\begin{table}[t]
\centering
\caption{Configured shot budget for each Clifford layout. Counts are the
same for the reference and checked layouts. A setting is a circuit and
measurement-basis template before randomization. The total includes
characterization; only the first row contributes target data to
Fig.~\ref{fig:clifford}\textbf{b}.}
\label{tab:clifford-budget}
\begin{ruledtabular}
\begin{tabular}{l r r r r}
Circuit family & Settings & Randomizations per setting & Shots per randomization & Shots per layout\\
\hline
Target stabilizers & 3 & 500 & 100 & 150\,000\\
Repeated-layer characterization & 54 & 500 & 100 & 2\,700\,000\\
SPAM reference & 1 & 500 & 100 & 50\,000\\
Auxiliary graph-state characterization & 16 & 500 & 100 & 800\,000\\
\hline
Total & 74 & & & 3\,700\,000\\
\end{tabular}
\end{ruledtabular}
\end{table}

\subsection{Physical noise characterization}
\label{app:clifford-learning}

We characterize the two CZ layer types separately. For each type, nine local Pauli preparation/measurement settings are repeated at layer depths $2,4,16$, accounting for $2\times9\times3=54$ characterization settings. The local basis covers single-qubit Paulis and weight-two Paulis on the ring edges. A separate depth-zero reference characterizes state preparation and measurement (SPAM). Repeated-layer expectation values are fitted to an exponential decay with a SPAM prefactor, and the decay factors are used to fit the Pauli-Lindblad rates~\cite{bergProbabilisticErrorCancellation2023}.

The model used for the reported predictions is a joint fit that also incorporates three auxiliary 7-regular graph states for each layout, inspired by the ACES protocol~\cite{flammia2021averaged} and recently utilized in Ref.~\cite{barron2026observable}. They use all $N$ qubits as data, including the four qubits that serve as checks in the checked target preparation. Each auxiliary graph has $N$ canonical weight-eight stabilizers, giving 60 and 72 additional observable constraints for the reference and checked layouts. Grouping these observables requires 16 settings in each case. The auxiliary preparations share the target's two unique CZ layers, while their different single-qubit Clifford sequences give additional constraints on the noise parameters. We emphasize that the target stabilizers in Eq.~\eqref{eq:clifford-groups} are not used as observations in this fit.

For each layer type the fitted support contains $3N$ single-qubit Pauli terms and $9N$ weight-two terms on the $N$ ring edges. The two layer types therefore have 480 fitted coefficients for the reference and 576 for the checked layout. The joint least-squares system combines the Pauli fidelities learned from the repeated layers with auxiliary-graph constraints of the form $-\log(F)/2=\sum_\nu c_\nu\lambda_\nu$, where $c_\nu$ counts anticommutations along the propagated observable and $F$ is the observed value after mitigating SPAM errors.

To assess the quality of the checks, we classify by syndrome the 6912 single-qubit and nearest-neighbor two-qubit Pauli faults in $V$ for the checked circuit using Eq.~\eqref{eq:syndrome-commutation}. Of these, 4475 have nonzero syndrome, giving the 65\% detection fraction reported in the main text.

\subsection{Rescaled estimates and uncertainty}
\label{app:clifford-analysis}
We apply the rescaling of Eq.~\eqref{eq:linearized-trace} to estimate the observables. For a fixed observable, let $S_r$ be its sum of post-selected outcomes and $N_r^{\rm ps}$ the post-selected count in randomization $r$. Each randomization receives $n_{\rm shot}=100$ shots. Define $\widehat\alpha_r=N_r^{\rm ps}/n_{\rm shot}$ and, when $N_r^{\rm ps}>0$, $\widehat m_r=S_r/N_r^{\rm ps}$. Multiplying $\widehat m_r$ by $\widehat\alpha_r$ restores the outcome sum per shot; dividing by $\overline{\widehat\alpha}$ supplies the common normalization. With the number of randomizations $R$,
\begin{equation}
    \widehat m
    =\frac1R\sum_{r=1}^R
    \frac{\widehat\alpha_r}{\overline{\widehat\alpha}}\widehat m_r
    =\frac{\sum_r S_r}{\sum_r N_r^{\rm ps}},
    \qquad
    \overline{\widehat\alpha}=\frac1R\sum_r\widehat\alpha_r.
    \label{eq:clifford-rescaling}
\end{equation}
For $N_r^{\rm ps}=0$, the product $\widehat\alpha_r\widehat m_r$ is defined as zero; the rescaled estimate requires a nonzero total post-selected count.

The rescaled terms share the estimated normalization $\overline{\widehat\alpha}$, so their ordinary standard error would ignore its fluctuations and its covariance with the numerator. First-order propagation of both gives the centered contributions
\begin{equation}
    \widehat u_r
    =\frac{\widehat\alpha_r}{\overline{\widehat\alpha}}
    (\widehat m_r-\widehat m)
    =\frac{S_r-\widehat m N_r^{\rm ps}}{\overline{N^{\rm ps}}},
    \qquad \overline{N^{\rm ps}}=\frac1R\sum_r N_r^{\rm ps}.
\end{equation}
The count expression defines $\widehat u_r=0$ when $N_r^{\rm ps}=0$. The estimated asymptotic standard error is therefore
\begin{equation}
    \widehat{\operatorname{SE}}(\widehat m)
    =\sqrt{\frac{1}{R(R-1)}
      \sum_{r=1}^R
      \left(
        \frac{S_r-\widehat m N_r^{\rm ps}}
             {\overline{N^{\rm ps}}}
      \right)^2}.
    \label{eq:clifford-sem}
\end{equation}
The factor $R-1$ is the sample-variance correction. If each randomization is reused for multiple shots rather than sampled independently for each shot, there can be a finite-sample ratio bias. The rescaled estimator then consistently estimates Eq.~\eqref{eq:linearized-trace}, but is not guaranteed to be exactly unbiased at finite sample size. Without post-selection, $N_r^{\rm ps}=100$ and the formula reduces to the ordinary standard error of the mean.

\subsection{Validation of the perturbative expansion}
For Fig.~\ref{fig:clifford}\textbf{c}, the checked model is evaluated at ten logarithmically spaced rate scales from $c=0.1$ to $c=2$. Each scale uses $10^9$ Monte Carlo samples of the modeled faults. As Pauli-Lindblad rates can be negative, we use a quasiprobability sampling strategy where needed~\cite{seif2026single}. We then compare these results with first- and second-order truncations. When scaling error rates, generator rates corresponding to check SPAM errors are scaled with $c$, while the data SPAM errors are held fixed and applied equally to the Monte Carlo and truncated predictions. The point at $c=1$ gives an estimate of the bias introduced by the second-order approximation in the experiment. The plotted error is the mean absolute discrepancy over the 20 stabilizers. Its experimental reference band is twice the mean per-observable standard error of the post-selected target data.

\section{Details on non-Clifford Ising dynamics}
\label{app:ising-details}

\subsection{Circuit setup and shot budget}

The non-Clifford circuit experiment is performed through a cloud-based submission of quantum circuits to the IBM Quantum 156-qubit Heron r3 device \textit{ibm\_aachen} using qiskit~\cite{qiskit2024}. The native gate set of the device to which all circuits are transpiled is $\{\text{CZ},\sqrt{X},R_z\}$ on a heavy-hexagonal topology.  At the time of performing the non-Clifford experiments, the device reported median values of $T_1 = 219\,\upmu \text{s}$, $T_2 = 250\,\upmu \text{s}$, average gate infidelities of $f_\text{CZ} = 1.63\times 10^{-3}$ and $f_{\sqrt{X}} = 2.08\times 10^{-4}$, and median readout errors of $0.49\,\%$. The 49-qubit subset of the device-layout shown in Fig.~\ref{fig:ising_ed_pec}\textbf{a} was chosen among the possible configurations such that the product of the measurement fidelities of the syndrome qubits was optimized. 
 
In table~\ref{tab:circuit-params}, we summarize the gate counts and shot budget for each depth setting. The circuit randomizations include both Pauli twirling of the CZ gate layers and measurements as well as the sampling of PEC terms. For the more shot-intensive techniques of full PEC and spacetime PEC, we use 160 shots for each of the $\mathcal{O}(10\text{k})$ circuit randomizations. The circuit randomizations are implemented through the \emph{Samplomatic} package~\cite{samplomatic}, which automates the generation of the $R_z$ gate angles that correspond to the sampled Pauli twirling and PEC-injected Pauli terms, given a noise model in sparse Pauli-Lindblad format of Eq.~\eqref{eq:phys-lind}. This enables sampling rates of around $3.5\,\text{kHz}$. For the more lightweight twirling-only runs used to obtain the unmitigated and error-detection-only baselines, we use 1024 circuit randomizations with 25 shots each. This is reflected in the error bars in panels \textbf{c}/\textbf{d} of Fig.~\ref{fig:ising_ed_pec} and in Figs.~\ref{fig:one-body_full} and~\ref{fig:two-body_full}, where the error bars of the ED and ED+PEC data points are of similar size, despite the sampling overhead incurred by the ED+PEC curve.

\begin{table}[b]
\centering
\caption{Circuit parameters and shot budget at each Trotter depth.}
\label{tab:circuit-params}
\begin{minipage}{0.8\textwidth}
\begin{ruledtabular}
\begin{tabular}{c c c c c c c}
 \begin{tabular}{@{}c@{}}Trotter \\ steps $d$ \end{tabular}& 
\begin{tabular}{@{}c@{}}CZ\\gates\end{tabular} & 
\begin{tabular}{@{}c@{}}$R_z$\\gates\end{tabular}& 
\begin{tabular}{@{}c@{}}$\sqrt{X}$\\gates\end{tabular}&
\begin{tabular}{@{}c@{}}Randomizations\\(PEC and ED+PEC)\end{tabular} &
\begin{tabular}{@{}c@{}}Shots\\(PEC and ED+PEC)\end{tabular} & \begin{tabular}{@{}c@{}}Sampling rate\\ in Hz \end{tabular} \\
\hline
2 & 216 & 446  & 272 & 10\,000 & 160 &   3571\\
4 & 432 & 740  & 468 & 20\,000 & 160 &   3560\\
6 & 648 & 1034 & 664  & 40\,000 & 160 &  3544\\
\end{tabular}
\end{ruledtabular}
\end{minipage}
\end{table}

\subsection{Post-selection protocol}
\label{app:HC_postselection}
 
As mentioned in the main text, two post-selection stages are applied sequentially.
First, all experimental runs, including noise learning circuits and SPAM error rescaling references, are subject to a post-selection procedure that is tailored to capture non-Markovian errors that would break the Pauli error channel formalism. Specifically, after the terminal measurement, each qubit is re-measured after a slow $R_x(\pi)$ pulse calibrated to have low leakage. A shot is discarded if any qubit fails to flip correctly between the two measurements under this pulse, see Ref.~\cite{barron2026observable} for more details. Typical survival rates of this procedure range from 16\% to 24\% depending on depth and circuit type, see Table~\ref{tab:survival-rates}. Afterwards, for those runs that rely on syndrome checks (the ED and ED+PEC curves), every shot in which any syndrome qubit did not return to $\ket{0}$ are discarded. Table~\ref{tab:survival-rates} reports the survival rates for each depth and circuit type. Let $H$ denote passing the non-Markovian-error filter and $S$ the trivial syndrome. The main-text post-selection and overhead convention use
\begin{equation}
    \alpha=\Pr(S\mid H)=\frac{\Pr(S\cap H)}{\Pr(H)},
    \qquad \GammaEDPEC=\frac{\gamma_L^2}{\alpha},
\end{equation}
where $\gamma_L$ is the quasiprobability norm of the first-order inverse. At $d=6$, the ED+PEC entries in Table~\ref{tab:survival-rates} give $\Pr(S\cap H)\approx0.032$ and $\alpha\approx0.197$. Thus the overall survival probability from all attempted shots is distinct from $\alpha$. The overheads in Table~\ref{tab:overhead} exclude the common filter's cost. Including that cost would multiply each method's overhead by the inverse of its own $\Pr(H)$.

The overall survival probability need not equal the product of the two marginal survival probabilities $\Pr(H)\Pr(S)$, because the filters are correlated; it always equals $\Pr(H)\Pr(S\mid H)$. A leakage event on a check qubit can be rejected by both stages. As expected, the measured syndrome post-selection rates with and without first-order PEC are similar. Within the deterministic Pauli model, zero-syndrome Pauli branches preserve the post-selection probability exactly, provided their insertion leaves the noise unchanged, as specified in Appendix~\ref{app:pec-estimator}.

\begin{table}
\centering
\caption{%
    Survival rates for each depth and circuit type. \textit{Non-Markovian} is the fraction of shots surviving a consistency check of successful measurement flips after the terminal measurement. \textit{Syndrome} is the fraction of overall shots (without the non-Markovian post-selection) surviving syndrome post-selection.
    \textit{Combined} is the fraction surviving both stages.
    Vanilla PEC uses only non-Markovian post-selection.
}
\label{tab:survival-rates}
\begin{minipage}{0.8\textwidth}
\begin{ruledtabular}
\begin{tabular}{c l c c c}
$d$ & Type & Non-Markovian errors & Syndrome & Combined \\
\hline
2 & Unmitigated & 0.242 & 0.364 & 0.132 \\
2 & PEC         & 0.245 & ---   & ---   \\
2 & ED+PEC      & 0.241 & 0.358 & 0.129 \\
\hline
4 & Unmitigated & 0.194 & 0.186 & 0.063 \\
4 & PEC         & 0.206 & ---   & ---   \\
4 & ED+PEC      & 0.209 & 0.194 & 0.071 \\
\hline
6 & Unmitigated & 0.172 & 0.103 & 0.034 \\
6 & PEC         & 0.162 & ---   & ---   \\
6 & ED+PEC      & 0.162 & 0.099 & 0.032 \\
\end{tabular}
\end{ruledtabular}
\end{minipage}
\end{table}
 
\subsection{Noise model validation}
\label{app:syndrome_rate_consistency_check}

As a consistency check of the learned Pauli noise models and the validity of the first-order approximation  of the logical noise channel, we compare the syndrome post-selection rates predicted by the noise model (to first order) with those actually measured in the experiments. In the first-order approximation $\LL=\LL^{(1)}$, we neglect the combinations of two or more detectable faults with zero total syndrome. The probability of measuring a trivial syndrome is therefore, to first order, equal to the probability that no detectable faults occur in the circuit. We can directly compute this from the generator rates $\lambda_\nu$, as each Pauli fault occurs independently with probability $(1 - e^{-2 \lambda_{\nu}})/2$. We further model readout noise as a symmetric bit-flip noise channel and extract the bit-flip rate from twirled depth-0 circuits. We compare this predicted post-selection rate to the measured one, obtained by the ratio of trivial-syndrome shots among the ones that pass the non-Markovian error post-selection step as outlined above.

Table~\ref{tab:model-validation} compares the predicted post-selection rate with the measured fraction of trivial-syndrome shots among those passing the non-Markovian-error filter. We find that the predicted and measured rates match closely. This validates the noise learning step and confirms that the first-order approximation of the spacetime fault classification is a reasonable one for the chosen circuit volumes, but does not by itself bound observable bias from combinations of detected faults that survive post-selection, which we quantify to second order in Appendix~\ref{app:second_order_bound}.

\begin{table}
\centering
\caption{%
    Predicted versus measured trivial-syndrome post-selection rates $\alpha$ (conditioned on passing the non-Markovian error filter).
}
\label{tab:model-validation}
\begin{minipage}{0.4\textwidth}
\begin{ruledtabular}
\begin{tabular}{@{}ccc@{}}
$d$ &
\begin{tabular}{@{}c@{}} Predicted\\ \end{tabular} &
\begin{tabular}{@{}c@{}} Measured\\ \end{tabular} \\
\hline
2 & 0.547 & 0.534 \\
4 & 0.339 & 0.339 \\
6 & 0.208 & 0.197 \\
\end{tabular}
\end{ruledtabular}
\end{minipage}
\end{table}
 
\subsection{Full single-site and nearest-neighbor two-body magnetizations}
Here we present the full experimental data that supports Fig.~\ref{fig:ising_ed_pec}, as well as additional data on the mitigation of two-qubit observables. All estimators and error bars are computed as outlined in Appendix~\ref{app:clifford-analysis}. In the main text, we show the average single-site magnetization for Trotter depths $d \in \{2, 4, 6\}$ in Fig.~\ref{fig:ising_ed_pec} and the full curve of the single-site magnetizations in Fig.~\ref{fig:ising_ed_pec}\textbf{d}. The corresponding curves for the single-site magnetizations at all depths are shown in Fig.~\ref{fig:one-body_full}. We observe the same qualitative behavior for all depths $d$, where ED+PEC recovers agrees best with the theoretical prediction among all methods. For $d=2$ the mitigation overhead of full PEC is small enough that it also recovers the theory well. 

From the same set of collected hardware samples, we also estimate the nearest-neighbor two-body magnetization $\langle X_i X_j\rangle$ for all 27 edges $(i,j)$ of the data graph at each Trotter depth, see Fig.~\ref{fig:two-body_full}. The qualitative conclusions mirror those from the single-site magnetization in the main text: ED+PEC accurately recovers the ideal values, while standard PEC degrades rapidly due to a prohibitive sampling overhead, whereas error detection alone provides only partial improvement. The ideal expectation values are somewhat lower in magnitude than the single-site case, reflecting the higher observable weight.

\begin{figure}[p]
    \centering
    \includegraphics[width=\columnwidth]{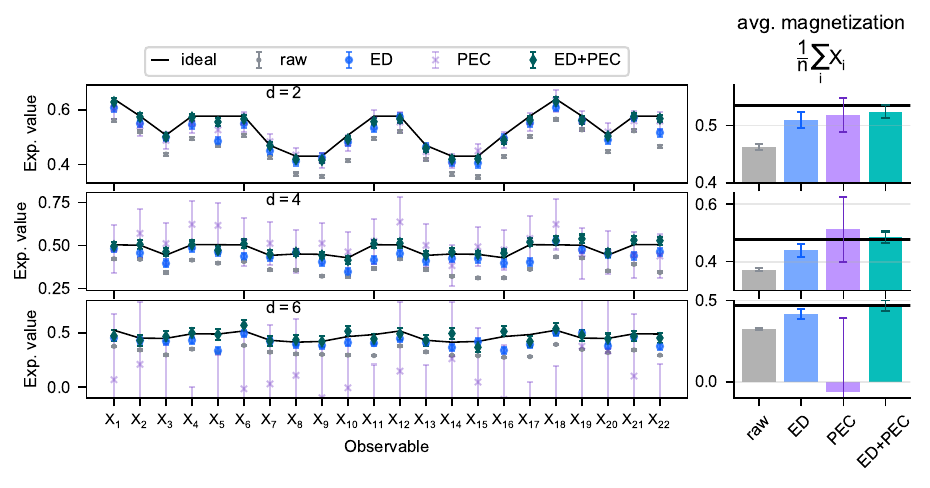}
    \caption{%
    Full data of the single-site magnetizations $\langle X_i\rangle$ for all 22 data qubits at Trotter depths $d=2$ (top), $d=4$ (middle), and $d=6$ (bottom) complementing Fig~\ref{fig:ising_ed_pec} of the main text. 
    }
    \label{fig:one-body_full}
\end{figure}

\begin{figure}[p]
    \centering
    \includegraphics[width=\columnwidth]{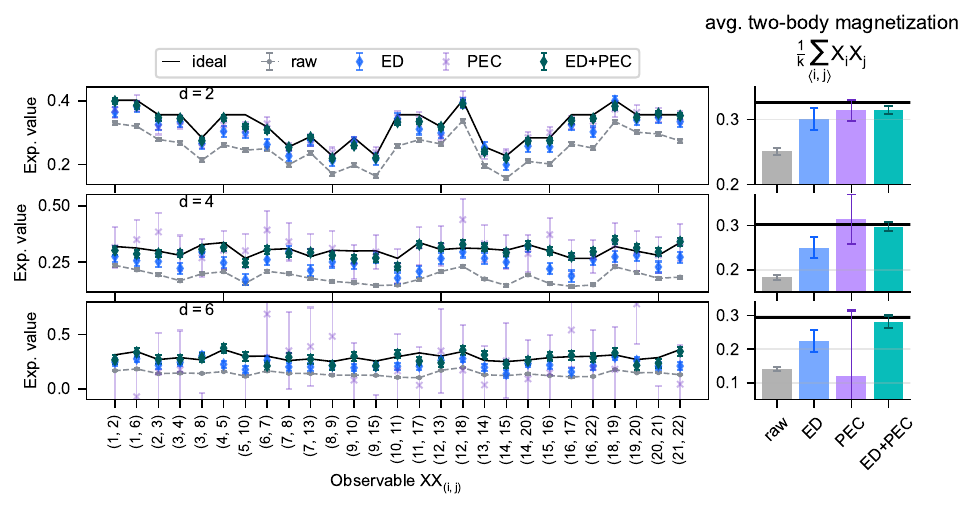}
    \caption{%
        Nearest-neighbor two-body magnetization $\langle X_iX_j\rangle$
        for all 27 bonds of the six-plaquette data graph at Trotter
        depths $d=2$ (top), $d=4$ (middle), and $d=6$ (bottom).
        Layout and conventions as in Fig.~\ref{fig:ising_ed_pec}.
    }
    \label{fig:two-body_full}
\end{figure}
 
\subsection{Systematic error of the first-order truncation}
\label{app:second_order_bound}

Our mitigation protocol inverts only the undetected part of the logical  generator. Here we bound the observable bias due to second-order terms, which is the leading contribution to the bias of this truncation. We first relate the omitted generator to observable bias and show that, within the convergence regime and at fixed circuit structure, retaining terms through order $m$ gives a bias of $O(c^{m+1})$ when noise rates scale as $\lambda_\nu\mapsto c\lambda_\nu$. We then specialize to the experimentally realized case of first-order PEC using the logical channel from Appendix~\ref{app:higher-order}.

Let $\LL^{(\leq m)}$ be the generator retained through homogeneous degree $m$ in the physical rates. Since all auxiliary Pauli conjugation maps commute, the modeled residual after applying its PEC inverse is
\begin{equation}
    \mathfrak R_m=e^{-\LL^{(\leq m)}}e^{\LL}
       =e^{\LL-\LL^{(\leq m)}}.
    \label{eq:bound-residual}
\end{equation}
To relate this auxiliary map to a measured error, write an auxiliary Pauli-map expansion $\mathfrak T=\sum_Q c_Q\mathfrak P_Q$ and define
\begin{equation}
    \|\mathfrak T\|_{\rm P,1}=\sum_Q|c_Q|.
    \label{eq:bound-coefficient-norm}
\end{equation}
Equal Pauli trajectories are combined before taking the norm. For $Q=\mathbb Q_{\bm a}$, the corresponding physical circuit $\mathcal C_{\rm phys}$ has the operator $K_{\bm a}=P_{a_M}U_M\cdots P_{a_1}U_1$ introduced in the main text. Define the expectation for this trajectory by
\begin{equation}
    O_Q=\operatorname{Tr}\!\left[O K_{\bm a}\rho K_{\bm a}^\dagger\right].
    \label{eq:bound-trajectory-expectation}
\end{equation}
An observable on the data register has an implicit identity on the checks. Since $K_{\bm a}$ is unitary and $\rho$ is normalized, $|O_Q|\leq1$ (assuming a Pauli observable $O$). The identity trajectory has $K_{\bm0}=U$ and
$O_I=\langle O\rangle_{\rm ideal}$.

Therefore, an expansion $\mathfrak R_m=\sum_Q r_Q\mathfrak P_Q$ gives $\langle O\rangle_{\rm PEC,\leq m}=\sum_Q r_Q O_Q$. Thus, writing $\mathfrak R_m-\mathfrak I=\sum_Q c_Q\mathfrak P_Q$, linearity in trajectory weights implies
\begin{equation}
\begin{aligned}
    |\Delta_m O|
    &\equiv|\langle O\rangle_{\rm PEC,\leq m}
                      -\langle O\rangle_{\rm ideal}|\\
    &=\left|\sum_Q c_Q O_Q\right|
      \leq\sum_Q|c_Q|
      =\|\mathfrak R_m-\mathfrak I\|_{\rm P,1}.
\end{aligned}
    \label{eq:bound-observable}
\end{equation}
Here the PEC expectation denotes the limit of the estimator as the number of shots tends to infinity. This argument does not require the residual map itself to have positive coefficients, as in the norm bounds in Ref.~\cite{govia2024bounding}.

The coefficient norm is submultiplicative and composing two Pauli-map sums multiplies their coefficients and can only decrease their absolute sum when equal trajectories are combined. Hence $\|\mathfrak T\mathfrak S\|_{\rm P,1} \leq\|\mathfrak T\|_{\rm P,1}\|\mathfrak S\|_{\rm P,1}$. Expanding the exponential in Eq.~\eqref{eq:bound-residual} gives
\begin{equation}
    |\Delta_m O|\leq e^{\varepsilon_m}-1
    \quad\hbox{whenever}\quad
    \|\LL-\LL^{(\leq m)}\|_{\rm P,1}\leq\varepsilon_m.
    \label{eq:bound-exponential}
\end{equation}
When scaling $\lambda_\nu\mapsto c\lambda_\nu$ at fixed circuit structure, the omitted generator begins at order $c^{m+1}$. Within the convergence regime, one may therefore take $\varepsilon_m=O(c^{m+1})$. Since $e^{\varepsilon_m}-1=\varepsilon_m+O(\varepsilon_m^2)$, this implies $|\Delta_m O|=O(c^{m+1})$ as expected from our perturbative expansion.

We now specialize to first-order PEC. In this case, the exact factorization in Appendix~\ref{app:higher-order} lets us bound the residual directly, without bounding the omitted generator. First-order PEC removes $\LL^{(1)}=\Lundet$ exactly. Recall from Appendix~\ref{app:higher-order} that $t_\nu=\tanh\lambda_\nu$, $t_A=\prod_{\nu\in A}t_\nu$, and $\bm s(A)=\bigoplus_{\nu\in A}\bm s_\nu$. The residual is
\begin{equation}
    \mathfrak R_1=\frac{\mathfrak B_{\rm d}}{z_{\rm d}},\qquad
    \mathfrak B_{\rm d}=\sum_{\substack{A\subseteq\Vdet\\\bm s(A)=\bm0}}
       t_A\mathfrak P_A,\qquad
    z_{\rm d}=\sum_{\substack{A\subseteq\Vdet\\\bm s(A)=\bm0}}t_A.
    \label{eq:bound-first-residual}
\end{equation}
The empty configuration has weight one. Subtracting the identity gives
\begin{equation}
    \mathfrak R_1-\mathfrak I
       =\frac1{z_{\rm d}}
       \sum_{\substack{\varnothing\ne A\subseteq\Vdet\\\bm s(A)=\bm0}}
          t_A(\mathfrak P_A-\mathfrak I).
\end{equation}
Define the sum of absolute weights of nonempty zero-syndrome configurations of detected faults as
\begin{equation}
    b=\sum_{\substack{\varnothing\ne A\subseteq\Vdet\\\bm s(A)=\bm0}}
          |t_A|.
    \label{eq:bound-accepted-weights}
\end{equation}
Since $\|\mathfrak P_A-\mathfrak I\|_{\rm P,1}\leq2$, Eq.~\eqref{eq:bound-observable} implies
\begin{equation}
    |\Delta_1 O|\leq\frac{2b}{|z_{\rm d}|}.
    \label{eq:bound-first-general}
\end{equation}

Evaluating Eq.~\eqref{eq:bound-first-general} directly requires the sums $b$ and $z_{\rm d}$ over zero-syndrome configurations. Their number can grow exponentially with the number of elementary faults. We therefore bound the contribution of the second order in $b$ by keeping the two-fault configurations with zero syndrome explicitly and discarding all configurations with three or more detected faults.

All sums and products in the following run over $\nu\in\Vdet$. Two detected faults have zero total syndrome precisely when their syndromes agree, so their contribution to $b$ is
\begin{equation}
    B_2=\sum_{\substack{\nu<\mu\\\bm s_\nu=\bm s_\mu}}|t_\nu t_\mu|, \quad \text{with} \quad  b = B_2 + O(c^3).
    \label{eq:bound-weight-upper}
\end{equation}

We also need a lower bound on $z_{\rm d}$. Using $\delta_{\bm s,\bm0}=2^{-k}\sum_{\bm x\in\{0,1\}^k} (-1)^{\bm x\cdot\bm s}$ to select zero syndrome in Eq.~\eqref{eq:bound-first-residual}, and $1\pm\tanh\lambda=e^{\pm\lambda}/\cosh\lambda$, gives
\begin{equation}
    z_{\rm d}
       =\frac{2^{-k}}{\prod_\nu\cosh\lambda_\nu}
         \sum_{\bm x\in\{0,1\}^k}
         \exp\!\left[\sum_{\nu\in\Vdet}
                   \lambda_\nu(-1)^{\bm x\cdot\bm s_\nu}\right]
       \geq\frac1{\prod_\nu\cosh\lambda_\nu}>0.
    \label{eq:bound-normalization-lower}
\end{equation}
For every detected fault, $\bm s_\nu\ne\bm0$, so $(-1)^{\bm x\cdot\bm s_\nu}$ averages to zero over $\bm x$. The average exponent is therefore zero, and convexity of the exponential implies that its average is at least one. This argument holds for either sign of every rate. Combining Eqs.~\eqref{eq:bound-first-general},
\eqref{eq:bound-weight-upper}, and~\eqref{eq:bound-normalization-lower}
yields
\begin{equation}
    |\Delta_1 O|\leq
       2\left(\prod_{\nu\in\Vdet}\cosh\lambda_\nu\right)
       \bigl(B_2+O(c^3) \bigr).
    \label{eq:bound-first-simple}
\end{equation}

We evaluate this bound to second order in the scaled generator rates for the observables $X_i$ shown in Figs.~\ref{fig:ising_ed_pec} and~\ref{fig:one-body_full}, and present the results in  Table~\ref{tab:second-order-bound}. We make the bound slightly tighter by omitting fault pairs where both spacetime Pauli faults lie outside of the causal lightcone of the observable from the sum entering $B_2$ in Eq.~\eqref{eq:bound-weight-upper}, as they do not contribute to the bias. We find that the second-order contributions remain smaller than the statistical error of the experimentally obtained values.  

\begin{table}
\centering
\caption{%
    Bounds on the systematic error from neglecting the second-order spacetime PEC contribution for single-qubit observables $X_i$ (averaged over all 22 data qubits). 
}
\label{tab:second-order-bound}
\begin{minipage}{0.6\textwidth}
\begin{ruledtabular}
\begin{tabular}{@{}ccc@{}}
$d$ &
\begin{tabular}{@{}c@{}} number of second order terms \\ \end{tabular} &
\begin{tabular}{@{}c@{}} Systematic error bound \\ \end{tabular} \\
\hline
2 & 126\,300 & 0.0014 \\
4 & 956\,521 & 0.0130 \\
6 & 2\,459\,257 & 0.0355 \\
\end{tabular}
\end{ruledtabular}
\end{minipage}
\end{table}

\end{widetext}

\onecolumngrid

\end{document}